\documentclass[usenatbib,usedcolumn]{mnras}

\usepackage{graphicx}	
\usepackage{amsmath}	
\usepackage{textcomp}  

\usepackage{color}

\newcolumntype{d}{D{.}{.}{-1}}

\newcommand{\ppdd}{$P$--$\dot{P}$ diagram}

\newcommand{\nuddot}{$\ddot{\nu}$}
\newcommand{\nudot}{$\dot{\nu}$}
\newcommand{\dnu}{$\Delta\nu$}
 
\newcommand{\beq}{\begin{equation}} 
\newcommand{\eeq}{\end{equation}}

\newcommand{\psra}{PSR~B1757$-$24}
\newcommand{\psrb}{PSR~B1800$-$21}
\newcommand{\psrc}{PSR~B1823$-$13}
\newcommand{\aaa}{B1757$-$24}
\newcommand{\bb}{B1800$-$21}
\newcommand{\cc}{B1823$-$13}

\newcommand{\cme}[1]{{\color{black} #1}}

\title[Long-term braking indices for glitching pulsars]
{New long-term braking index measurements for glitching pulsars using a glitch-template method}

\author[C.~M. Espinoza et al.]{   
C.~M. Espinoza,$^{1,2}$\thanks{E-mail:cristobal.espinoza.r@usach.cl}  
A.~G. Lyne,$^{3}$ 
B.~W. Stappers$^{3}$\\
$^{1}$Departamento de F\'isica, Universidad de Santiago de Chile, Avenida Ecuador 3493, Estaci\'on Central, Santiago, Chile. \\
$^{2}$Instituto de Astrof\'isica, Facultad de F\'isica, Pontificia Universidad Cat\'olica de Chile, Casilla 306, Santiago 22, Chile. \\
$^{3}$Jodrell Bank Centre for Astrophysics, School of Physics and Astronomy,
The University of Manchester, Manchester M13 9PL, UK. 
}

\date{Accepted XXX. Received YYY; in original form ZZZ}
\pubyear{2016}
\begin{document}
\label{firstpage}
\pagerange{\pageref{firstpage}--\pageref{lastpage}}
\maketitle

\begin{abstract}

Braking index measurements offer the opportunity to explore the processes affecting the long-term spin evolution of pulsars and possible evolutionary connections between the various pulsar populations.
For young pulsars the long-term trends are generally obscured by short term phenomena such as timing noise and the recoveries form large glitches.
Here we present a new method to overcome the latter and report on braking index measurements for the Vela-like pulsars PSR \bb\ and PSR \cc, an updated measurement for Vela and new estimates for four more glitching pulsars observed at Jodrell Bank Observatory.
The values of braking indices describe the long-term evolution of the pulsars across the \ppdd.
Despite some measurements being affected by considerable uncertainties, there is evidence for a common trend involving low braking indices ($n\leq2$) among young glitching pulsars. 
Such values introduce a new variant in the evolution of young pulsars and their relationship with other populations in the \ppdd.
Low braking indices also imply that these pulsars could be a few times older than their characteristic ages. 
We consider PSR \aaa\ and conclude that the pulsar could be old enough to be related to the supernova remnant G5.4$-$1.2. 
Between glitches, the short-term evolution of Vela-like pulsars is characterised by large inter-glitch braking indices $n_\text{ig}>10$.
We interpret both short and long term trends as signatures of the large glitch activity, and speculate that they are driven by short-term post-glitch re-coupling and a cumulative long-term decoupling of superfluid to the rotation of the star.
\end{abstract}




\section{Introduction}
\label{intro}
Pulsars are the most commonly observable manifestation of highly magnetised and rapidly rotating neutron stars. 
Electromagnetic radiation is generated in their magnetospheres and directed at us on every turn of the star owing to the misalignment of the magnetic and spin axes.
Their rotation can be precisely tracked by detection and timing of the electromagnetic pulses arriving at Earth.
The exquisite accuracy of pulsar timing at radio wavelengths has permitted numerous applications, from tests of the general theory of relativity to the possibility of detection of gravitational waves \citep{lbk+04,hd83,jhlm05}, and therefore a good comprehension of the rotational dynamics of neutron stars is fundamental to optimise their use as precise celestial clocks.

It was shortly after their discovery when the monotonic decrease of pulsars spin frequency ($\nu$) was detected and the first spin-down rates ($\dot{\nu}<0$) were measured \citep{dhs69,col69}.
This is the most prominent behaviour of pulsar rotation.
The rotational energy losses were modelled as electromagnetic radiation produced by a rotating, slanted and constant magnetic dipole attached to the star \citep{go69}.

If the neutron star is considered as a magnetic dipole attached to a solid sphere rotating in vacuum and no other braking mechanism is in action, then $\dot{\nu}\propto-\nu^3$ and a measurement of the spin-down rate offers direct information on the strength of the magnetic dipole.
Generally, assuming a stellar radius of 10\,km, a moment of inertia $I=10^{45}$\,g\,cm$^2$, and the magnetic dipole perpendicular to the rotation axis, the surface magnetic dipole strength at the neutron-star equator is estimated as  $B=3.2\times10^{19}\sqrt{P\dot{P}}$~G \citep[e.g.][]{ls90}, where $P=1/\nu$ and $\dot{P}=-\dot{\nu}/\nu^2$.
In the same context, assuming that the initial spin frequency was much larger than the current value and that no other braking mechanism has acted on the star, the age of the pulsar can be calculated as $\tau_c=-\nu/2\dot{\nu}$.
This quantity is known as the characteristic age and, given the above assumptions, its wide use as an age proxy must be taken with caution.

The supposition that a neutron star is a solid sphere with an attached dipole rotating in vacuum might turn out to be too simplistic.
\cme{The broad multi-wavelength emission of pulsars, from radio to gamma rays, indicates that they must be surrounded by a plasma in which particles are accelerated up to very high energies. }
Such processes could effectively remove angular momentum from the star and alter the normal spin-down caused by the underlying magnetic braking \citep{mnd85,hck99,klo+06}.
Furthermore, an important fraction of the star's interior is believed to be in the form of a neutron superfluid \citep{bpp69} and part of this superfluid may not be coupled to the spin-down of the outer crust. 
Proof of this are glitches: sudden and sporadic spin-up events that interrupt the normal spin-down of pulsars \citep{rm69,elsk11}.
Glitches are thought to be caused by angular momentum transfer from a more rapidly rotating superfluid component to the crust \citep{ai75,hm15}.
Coupling and decoupling between this superfluid and the crust can not only produce glitches but, in principle, could also introduce long-term variations on the effective moment of inertia to which the external torque is applied, thereby generating deviations from the normal dipole spin-down \citep{ha12}.
Finally, structural changes like evolution of the direction of the dipole magnetic field could also interfere with the normal spin-down of a neutron star \citep[e.g.][]{mmw+06,lgw+13}. 
The emission of gravitational waves could also contribute to energy losses but upper limits on the gravitational wave emission of the Crab and Vela pulsars indicate contributions of less than $1\%$ and $10\%$, respectively \citep{aaa+14}.

\subsection{The Braking Index}
In an effort to recognise the mechanisms affecting the long-term spin evolution of pulsars, the braking index ($n$) is introduced to quantify the spin-down process.
It is defined by relating the spin frequency and its first derivative through a power law:
\beq
\label{eq:powerLaw}
\dot{\nu}=-\kappa\nu^n  \quad ,
\eeq
where $\kappa$ is normally assumed to be constant with time. 
If the spin-down is driven only by a constant dipole electromagnetic braking then $n=3$ and $\kappa$ would depend  on the moment of inertia of the star and the strength and orientation of the magnetic dipole.
A time-dependent $\kappa$, corresponding to changes of the moment of inertia or the magnetic dipole, would produce braking indices different from $3$.
The braking index would also differ from $3$ if higher order multipoles or other magnetic field configurations played an important role in the spin-down of the star.
If the electromagnetic torque was not the dominant spin-down process, measurements of $n$ could shed light on the properties of the driving torque  \citep{br88,ysl+13}

A more precise calculation of the age of the pulsar is possible if we know the real braking index and the initial spin rate $\nu_0$ \citep[e.g.][]{mnd85}:
\begin{equation}
\label{eq:age}
\tau=\left\{ \begin{array}{ll}
	\frac{2\tau_c}{n-1}\left[1-\left(\frac{\nu}{\nu_0}\right)^{n-1}\right] & \textrm{for}  \quad n\neq1 \\
	2\tau_c\ln\left(\frac{\nu}{\nu_0}\right) & \textrm{for} \quad  n=1. \\
\end{array} \right.
\end{equation}
We note that the above assumes a constant braking index over the pulsar's life time.
The braking index can be determined from measurements of the first two derivatives of $\nu$, since time-differentiation of Eq. \ref{eq:powerLaw} gives
\begin{equation}
\label{eq:index}
n=\frac{\nu \ddot{\nu}}{\dot{\nu}^2} \quad .
\end{equation} 
\begin{table*}
\begin{minipage}{165mm} 
  \caption{Known braking indices between glitches ($n_\text{ig}$) and long-term braking indices ($n$). 
    Uncertainties (1-sigma) on the last quoted digit are shown between parentheses.  
    Characteristic ages ($\tau_c$) and reference values for $\nu$ and $\dot{\nu}$ are shown too.}
  \label{tbl:enes}
   {\normalsize 
  \begin{tabular}{lldrdddl}
  \hline
  \multicolumn{1}{l}{Name} & \multicolumn{1}{l}{J name} & \multicolumn{1}{c}{$\nu$} & \multicolumn{1}{c}{$\dot{\nu}$} 
& \multicolumn{1}{c}{$\tau_c$} & \multicolumn{1}{c}{$n_\text{ig}$} &  \multicolumn{1}{c}{$n$} & \multicolumn{1}{l}{Refs.$^a$}  \\
  \multicolumn{1}{l}{} & \multicolumn{1}{c}{} & \multicolumn{1}{c}{Hz} & \multicolumn{1}{c}{$10^{-15}$\,Hz\,s$^{-1}$} 
& \multicolumn{1}{c}{kyr} & \multicolumn{1}{c}{} &  \multicolumn{1}{c}{} & \multicolumn{1}{l}{}  \\
   \hline
   B0531$+$21 (Crab)   & J0534+2200   &	  29.946 & -377535 & 1.26    & 2.519(2)    & 2.342(1)  & (1)            \\ 
   J0537$-$6910  & J0537$-$6910 & 62.018 & -199374 & 4.93    & \sim20      & -1.2      & (2)         \\ 
   B0540$-$69    & J0540$-$6919 & 19.775 & -187272 & 1.67    & 2.13(1)     &  --       & (3)         \\ 
   B0833$-$45 (Vela)   & J0835$-$4510 & 11.200   & -15660  & 11.3    & 41.5(3)     & 1.7(2)    & {\sl this work} \\ 
   J1119$-$6127  & J1119$-$6127 & 2.4473 & -24050  & 1.61    & 2.684(2)^b  &  --       & (4)         \\ 
   J1208$-$6238  & J1208$-$6238 & 2.2697 & -16843  & 2.67    & 2.598(1)    &  --       & (5)         \\ 
   B1509$-$58    & J1513$-$5908 & 6.6115 & -66944  & 1.56    & 2.832(3)    &  --       & (6)         \\ 
   J1734$-$3333  & J1734$-$3333 & 0.8551 & -1667   & 8.13    & 0.9(2)      &  --       & (7)         \\ 
   B1800$-$21    & J1803$-$2137 & 7.4825 & -7528   & 15.8    & 25.9(4)     & 1.9(5)    & {\sl this work}         \\ 
   B1823$-$13    & J1826$-$1334 & 9.8549 & -7313   & 21.4    & 29.5(4)     & 2.2(6)    & {\sl this work}         \\ 
   J1833$-$1034  & J1833$-$1034 & 16.159 & -52750  & 4.85    & 1.857(1)    &  --       & (8)         \\ 
   J1846$-$0258  & J1846$-$0258 & 3.0621 & -66640  & 0.73    & 2.65(1)^c   &  --       & (9)         \\ 
\hline 
\end{tabular} } \\
{\sc Note.---} \cme{Not included in this compilation are braking index measurements or reported $n$ switches that are based on observations having short time spans ($<5$\,yr) compared to those used in the above measurements.} \\
$^a$References: 
(1)  \citet{ljg+15};
(2) \citet{aeka15} (also \cite{mmw+06});
(3) \citet{fak15};  
(4) \cite{wje11};
(5) \cite{cpw+16}
(6) \cite{lk11};
(7) \cite{elk+11};
(8) \cite{rgl12};
(9) \cite{lkg+07}. \\
$^b$A possible reduction of about 15\% is observed after a large glitch \citep{awe+15}. \\
$^c$Value was found to decrease to $n=2.19$ after a large glitch \citep{lnk+11,akb+15}. 
\end{minipage} 
\end{table*}

So far, there are braking index measurements available for only 10 pulsars (without considering the new ones presented here; Table~\ref{tbl:enes}).
All exhibit values $n<3$, indicating that the braking must be more complex than simple  magnetic braking due to a constant dipole.
%
The reason that only a few values of $n$ have been determined comes from the difficulties found when trying to measure the long-term second derivative of the spin frequency. 
In most cases, the effects of a \nuddot\ arising purely due to simple spin evolution are too small compared with the effects of other phenomena which also affect the rotation, such as glitches and timing noise.
Timing noise refers to the wandering of the pulsar rotational phase around the predictions of a simple slowdown model, observed in the data of most pulsars \citep{hlk10,sc10}.
It was shown by \citet{lhk+10} that such deviations from the model can be caused by rapid, quasi-periodic switches between two or more \nudot\ values present in the data.
Furthermore, their analysis showed robust correlation between the \nudot\ switches and observed changes between two or more pulse profile shapes, thereby suggesting a magnetospheric origin for timing noise.
Thus, the long-term \nuddot\ might be detectable only if the associated \nudot\ monotonic variation, in a given time, is larger or at least comparable to the amplitude of the timing noise.
In most pulsars, the rate at which \nudot\ changes is extremely low and the detection of the long-term \nuddot\ is practically impossible, even on the longest datasets available.
In fact, for a constant braking index, the expected $\ddot{\nu}\propto\dot{\nu}/\tau_c$ (Eq. \ref{eq:index}) decreases rapidly as pulsars age, because \nudot\ decreases and $\tau_c$ increases. 
This is why all measured braking indices have been obtained only for very young pulsars, with $\tau_c\la10$\,kyr (those in Table \ref{tbl:enes}).

In addition to timing noise, the presence of glitches can also complicate the measurement of \nuddot.
Young pulsars like the Vela pulsar ($\tau_c\geq10$\,kyr) exhibit the highest known glitch activities but the very young pulsars, like the Crab ($\tau_c\sim1$\,kyr), exhibit somewhat lower glitch activities \citep{elsk11}. 
Consequently, most known braking indices have been measured for very young pulsars, with relatively low (or zero) glitch activity (with the Vela pulsar being the only exception). 
In the majority of these cases, if glitches had interrupted the spin-down evolution, the measurements were performed between glitches, often obtaining consistent values \citep{lps93,fak15}.
However, recent results suggest that large glitches (or increased glitch activity) can modify the braking index measured between glitches ($n_\text{ig}$). 
One case is the Crab pulsar, the first pulsar to have its braking index measured.
It exhibited a stable spin-down evolution between glitches, for $25$\,yr, with $n_\text{ig}=2.5$ \citep{gro75c,lps93}. 
After that, the pulsar experienced a period of about $11$\,yr in which the glitch activity increased and the braking index, calculated between glitches, decreased to $n_\text{ig}\sim2.3$ \citep{ljg+15}. 
Other examples are PSRs J1846$-$0258 and J1119$-$6127.
The only two large glitches observed in these very young pulsars are about an order of magnitude larger than those seen in the Crab pulsar and introduced effects almost twice as large, characterised by $n_\text{ig}$ reductions of $17\%$ and $15\%$, respectively \citep{akb+15,awe+15}\footnote{In the case of PSR J1119$-$6127, \cite{awe+15} indicate that the change of \nuddot\ could be also interpreted as a slow exponential relaxation (with a time constant close to 2,300 days) following the over recovery of \nudot\ after the glitch.}.
The above suggests that the long-term trends (with time scales longer than the inter-glitch intervals) might only be revealed after a number of large glitches.

Another effect of the glitches comes in the form of persistent negative steps in spin-down rate, which accumulate over time and lower the long-term, mean value of \nuddot.
After a number of steps the braking index $n$, calculated via the long-term \nuddot, will be smaller than any value calculated between glitches (i.e. $n\leq n_\text{ig}$; see Table \ref{tbl:enes}).
The amplitude of this difference varies between pulsars and it could be proportional to the size of the glitches \citep{ljg+15}.
For B0540$-$69, which has shown only a couple of small glitches in $\sim16$\,yr, the effect is almost negligible \citep{fak15}.
For the Crab pulsar, with some middle size glitches, the effect is close to $6\%$ and gives a long-term braking index $n=2.342(1)$ \citep{ljg+15}.

\begin{figure*} 
 \includegraphics[width=13cm]{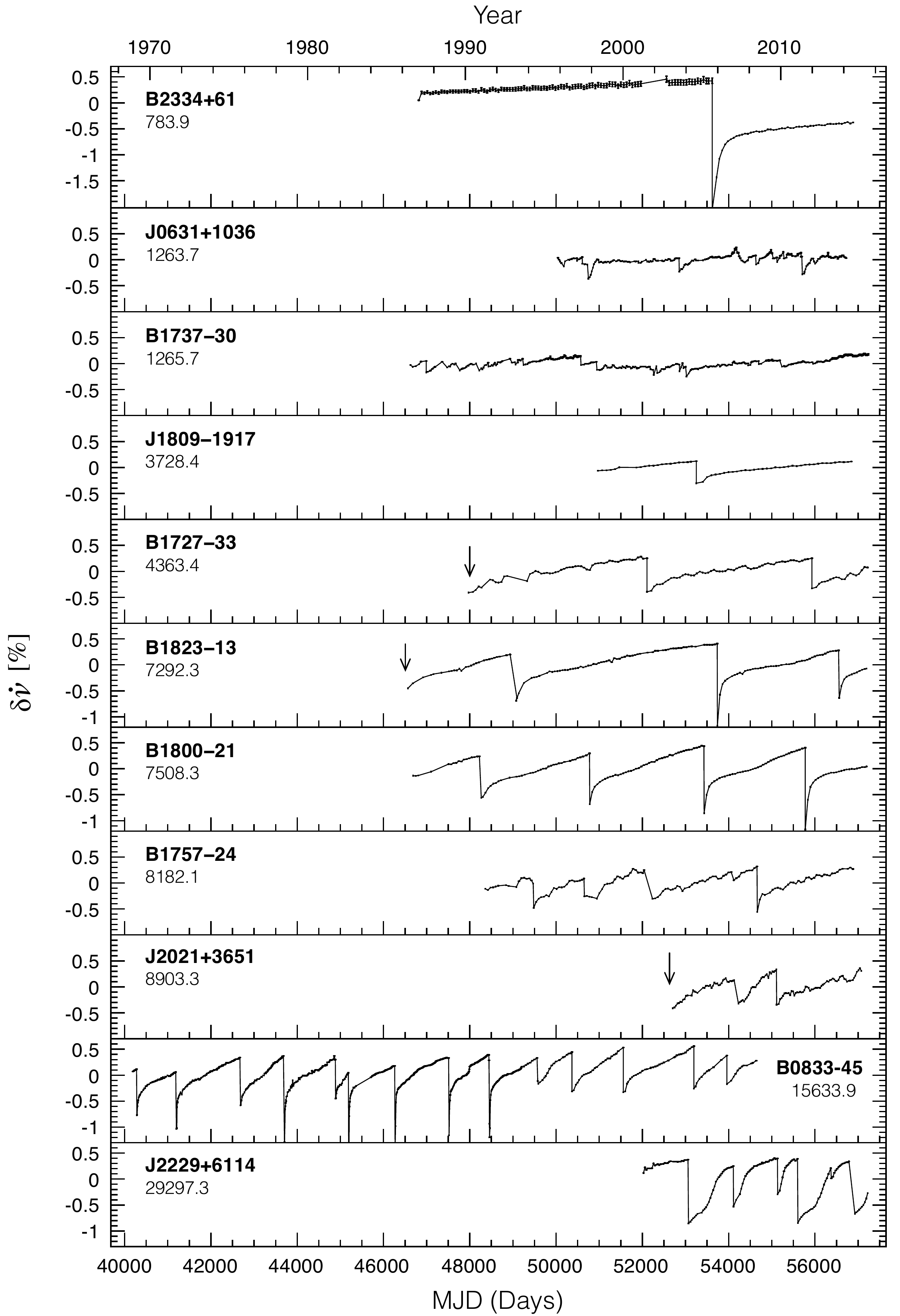} 
 \caption{The \nudot\ time-evolution of 11 young glitching pulsars. 
 For each pulsar, the percentage of the \nudot\ variation with respect to the mean observed value is plotted. 
 Both pulsar name and mean spin-down rate value $|\dot{\nu}|$ (in units of $10^{-15}$\,Hz\,s$^{-1}$) are indicated for each pulsar and the panels are sorted according to $|\dot{\nu}|$, increasing from top to bottom.
 All panels share the same \nudot\ scale. 
 Arrows indicate the epochs of glitches that occurred prior the start of the JBO observations.}
\label{allf1}
\end{figure*}

However, this effect becomes extreme for the young pulsars that exhibit the largest known glitches, like the Vela pulsar.
For these pulsars the evolution of \nudot\ is regularly interrupted by very large negative spin-down steps at the glitches; thereby complicating the detection of the long-term \nuddot\ (see the \nudot\ time evolution of Vela and other glitching pulsars in Fig. \ref{allf1}, where every step downward is a glitch).
Between glitches, however, the Vela pulsar evolves with braking indices $n_\text{ig}\sim40$ and  in some cases as high as $60$ \citep[][
also \S \ref{sec:largeglits} and Table \ref{interGs}]{ls90}.
In contrast, the long-term braking index of the Vela pulsar is significantly less than three, as shown by \citet{lpgc96}, who devised a method to overcome the presence of several large glitches and measure the braking index.

In this paper we develop their method and give new braking index measurements for the two Vela-like pulsars PSRs \bb\ and \cc\ plus a re-measurement for the Vela pulsar using an updated dataset. 
All these pulsars exhibit large and quasi regular glitches with strong effects that dominate their secular \nudot\ evolution.
They were selected because three or more of these glitches have already been detected in the  period they have been monitored.
In addition to these measurements, we also apply the method to other young glitching pulsars which show less regularity in their behaviour.
By doing so we find indications of a common behaviour among all these pulsars, characterised by low braking indices.
In section \ref{sct:velas} we comment on the main properties of the Vela-like pulsars relevant to this study.
In section \ref{curves} we describe the data used for the analyses and how \nudot\ datasets were constructed.
We also explain the method and, in the following sections, give and discuss our results.

\section{Vela-like pulsars}
\label{sct:velas}
The Vela pulsar (PSR B0833$-$45, J0835$-$4510) is the nearest young and energetic pulsar to Earth, and hence is the brightest pulsar in the radio sky.
It rotates 11 times a second and has a spin-down power ($\dot{E}=-4\pi^2I\nu\dot{\nu}$) of $7\times 10^{36}$\,ergs\,s$^{-1}$. 
A fraction of this energy is injected into the pulsar wind nebula (PWN) known as Vela-X,  which closely surrounds the pulsar \citep{wp80}.
The system is associated with the Vela supernova remnant (SNR), for which recent calculations suggest an age close to $\sim 10$\,kyr \citep[using temperature measurementes to calculate the expansion velocity,][]{sh14}.
This is similar to the pulsar's characteristic age $\tau_c=11$\,kyr, but we note that other estimates suggest SNR ages up to a factor of two larger \citep[e.g.][]{aet95,tsk+09,plps09}, consistent with the large age suggested by the small braking index \citep[Eq. \ref{eq:age};][]{lpgc96}.
Regardless of the uncertainties, all these estimates indicate Vela is a young pulsar, although not as young as the Crab and other pulsars which have ages closer to $1$\,kyr.

PSRs \bb\ and \cc\ share several characteristics with the Vela pulsar. 
They have similar rotational properties (with magnitudes of $\nu$ and \nudot\ just below those of Vela) and therefore have similar characteristic ages $\tau_c$ ($16$--$22$\,kyr), and about the same spin-down power of $2$--$3\times10^{36}$\,erg\,s$^{-1}$.
PSRs \bb\ and \cc\ are not clearly associated with any SNR \citep{bgl89,bckf06} but they both power X-ray PWNe, as expected with energetic pulsars \citep{fsp96,kpg07}.

Some of the above properties might not be exclusive to these three pulsars; actually, many other young pulsars are also associated with PWNs or SNRs but not considered Vela-like pulsars.
Their glitch properties, however, effectively separate these three pulsars (and a number of others) from the rest of the population.
\citet{elsk11} noted how Vela and other similar pulsars (their Table 6) exhibit the largest glitches among all pulsars and present a very low probability of experiencing small glitches.
Moreover, and contrary to what is observed in other pulsars, glitches in these pulsars seem to occur at quasi regular intervals of time \citep{lel99,mpw08}.
In this paper we use the term Vela-like to refer to pulsars with these glitching properties, which are described in more detail below. 

\subsection{Large glitches and their recoveries}
\label{sec:largeglits}
The Vela pulsar was the first pulsar found to glitch, in 1969 \citep{rm69}, and has been densely monitored by different radio observatories for the last 45 years. 
During this time its rotation has been regularly interrupted, on average every 3--4 years, by glitches of typical sizes $\Delta\nu\sim20\,\mu$Hz but on occasion as large as $35\,\mu$Hz (in general, $\Delta\nu/\nu>10^{-6}$), reaching the higher end of the observed \dnu\ distribution for all pulsars \citep{elsk11}.
The spin-down rate $\dot{\nu}$ also changes following the large glitches, becoming larger ($\Delta\dot{\nu}<0$) and relaxing back approximately exponentially with multiple time scales extending from minutes to a few hundred days \citep{sl96,dml02}.
Hence every large glitch is followed by a recovery period in which the rotation relaxes towards the pre-glitch configuration.
\cme{This could be a generic behaviour for all glitches but, due to detectability issues, it is unclear whether this extends to glitches of all sizes \citep[e.g.][]{eas+14}.}
In the case of Vela-like pulsars the spin-down step-changes at glitches are the largest observed \citep{elsk11} and the recoveries from these glitches have a dramatic effect on the long-term spin evolution. 
They completely dominate the time-evolution of \nudot\ between glitches, producing a significant increase of the inter-glitch \nuddot\ and leading to very large inter-glitch braking indices $n_\text{ig}$ (Fig. \ref{allf1}).

The influential imprints of the glitch recoveries make the \nudot\ curves of Vela-like pulsars look very similar.
During about $29$~yr of observations, PSR's \bb\ and \cc\ have exhibited four large glitches each. 
All of them have been followed by approximately exponential recoveries that imprinted repeating patterns in their $\dot{\nu}$ evolution, which resemble those of the Vela pulsar but with longer time scales, glitching every 5 to 7 years. 

Based on these properties, many other pulsars can also be regarded as Vela-like pulsars.
Visual inspection of the \nudot\ curves of these pulsars immediately suggests that they have similar glitching properties.
Examples of this are PSRs B2334+61, J1809$-$1917 and B1727$-33$ (Fig.~\ref{allf1}).
As it will be described later, our method needs the presence of at least three large glitches, making it impossible (at the moment) to include some of these pulsars further in this study, but they will continue to be monitored.

\section{Observations and \nudot\ evolution}
\label{curves}

Observations were carried out mostly with the 76-m Lovell telescope and some additional observations were performed with the 30-m MkII telescope (initially at 1600\,MHz but at 1400\,MHz after 1999), both at Jodrell Bank Observatory (JBO) in the UK.
The original 25 years of data for the Vela pulsar used in \citet{lpgc96} have been extended to nearly 40 years by observations with the 64-m telescope at Parkes.
Observations from the Parkes pulsar data archive were also used to complement the dataset of  PSR \bb.

With the Lovell telescope, pulsars were observed at typical intervals of 4 to 7 days in a 64-MHz band centred on 1404\,MHz using an analogue filter-bank.
Since mid-2009 observations were performed using a digital filter-bank backend using $1024\times0.5$\,MHz channels, of which approximately 380\,MHz is used. More details of the observation procedures can be found in \citet{hlk+04}.
In this paper we use data taken up to August 2015.

Pulse times of arrival (TOAs) are determined by convolving a standard pulse profile template with the observed profiles.
TOAs are corrected for the motion of the observatory around the Sun and compared with a simple slowdown model of the frequency and frequency derivative of the rotation of the pulsar to generate the so called timing residuals \citep[e.g][]{lk05}.
\cme{We use dispersion measure values and astrometry parameters as quoted in \citet{hlk+04}, 
which in some cases were mildly corrected using newer data and standard techniques.
The only exception is PSR B1727$-$30, for which we use an improved proper motion measurement provided by M. Keith (private communication). }
Minimisation of the timing residuals for a group of TOAs gives best estimates for the frequency and frequency derivative at a given epoch, normally chosen at the centre of the time-span defined by the TOAs.

We constructed \nudot\ curves using values obtained through fits of $\nu$, $\dot{\nu}$ and \nuddot\ to $120$-day-long groups of TOAs centred on epochs which were separated by $60$\,d.
The uncertainties of the \nudot\ values are the formal errors from the minimisation of the timing residuals. 
In order to reduce the gaps with no data at the glitches and to visually accentuate their sudden nature, extra \nudot\ values were calculated before and after the glitches, at the epochs of the last TOA before the glitch and the first TOA after the glitch.
The \nudot\ curves for all pulsars are presented in Fig. \ref{allf1}.
Glitch epochs were taken from Table~1 in \citet{elsk11} and from the JBO online glitch catalogue\footnote{\url{http://www.jb.man.ac.uk/pulsar/glitches.html}}. 

For the Vela pulsar, we use the \nudot\ data obtained by \citet{lpgc96} plus additional data points calculated from Parkes TOAs, from April 1992 to September 2008.
The former dataset was obtained from fits to groups of TOAs spanning $\sim10$ to 50 days, and separated by $\sim5$ to $\sim 30$ days. 
Because of their lower cadence, the \nudot\ data obtained from the Parkes TOAs were calculated from fits to groups of TOAs spanning 50 to up to about 200 days.
In both cases the sliding fit-window had a stride of about half the length of the window.

\section{Measuring braking indices in glitching pulsars}
In general, the measurement of $n_\text{ig}$ is straightforward, provided the rotation is not dominated by timing noise and the \nuddot\ signal is significant.
In most cases, standard timing techniques (i.e. the fit of the TOA rotational phases) will suffice and in other cases it is possible to use frequency data to perform the measurements \citep[as for the Crab pulsar and PSR B0540$-$69;][]{ljg+15,fak15}  or \nudot\ data, as we do for Vela and PSRs \bb\ and \cc\ in section \ref{sct:results}.

To measure the long-term braking index $n$, however, it is necessary to perform more detailed analyses in order to account for the glitches and timing noise correctly.
In general, standard timing techniques fail to detect the long-term \nuddot\ when the data includes a number of large glitches, such as those shown in Fig.~\ref{allf1}, and only report \nuddot\ values between the glitches.
A solution to this problem was proposed by \citet{lpgc96}, who looked at the overall long-term slope of the \nudot\ versus time relation. 
They noticed the repeatable nature of the glitch relaxation process and demonstrated, in the case of the Vela pulsar, that after the short term transients of a glitch the pulsar returns to a standard dynamical rotational configuration which may evolve following the secular long-term spin evolution.
Based on this assumption, they determined a braking index $n=1.4\pm 0.2$.
The method consisted of selecting points on the $\dot{\nu}$ curve a fixed number of days ($N_d$) after each glitch and then extrapolating from that point back to the glitch epoch, using the slope of the post-glitch curve from that point until the next glitch.
This process creates a second set of points that lie very close to a straight line, which is representative of the long-term $\dot{\nu}$ evolution, i.e $\ddot{\nu}$. 
Even though it was shown that the value of $n$ is not highly dependent on the chosen value for $N_d$, this somewhat arbitrary choice is clearly not optimal. 
Below we describe a new method, based on similar principles, which is essentially independent of such choices.

\subsection{The use of glitch \nudot\ templates}
\label{method}
We have developed a method to measure the braking index of pulsars that exhibit large and regular glitches taking advantage of the consistent relaxation patterns observed after each glitch.
The idea is to use a single template, for the post-glitch \nudot\ curves of a given pulsar, that can be used to measure the relative upward or downward shift of each individual post-glitch curve.
The series of shifts we get from fitting all the glitches for a given pulsar contain information on the general, long-term slope of \nudot\ (i.e. \nuddot) and can therefore lead to a long-term braking index measurement.

\begin{figure} \begin{center}
\includegraphics[width=8.3cm]{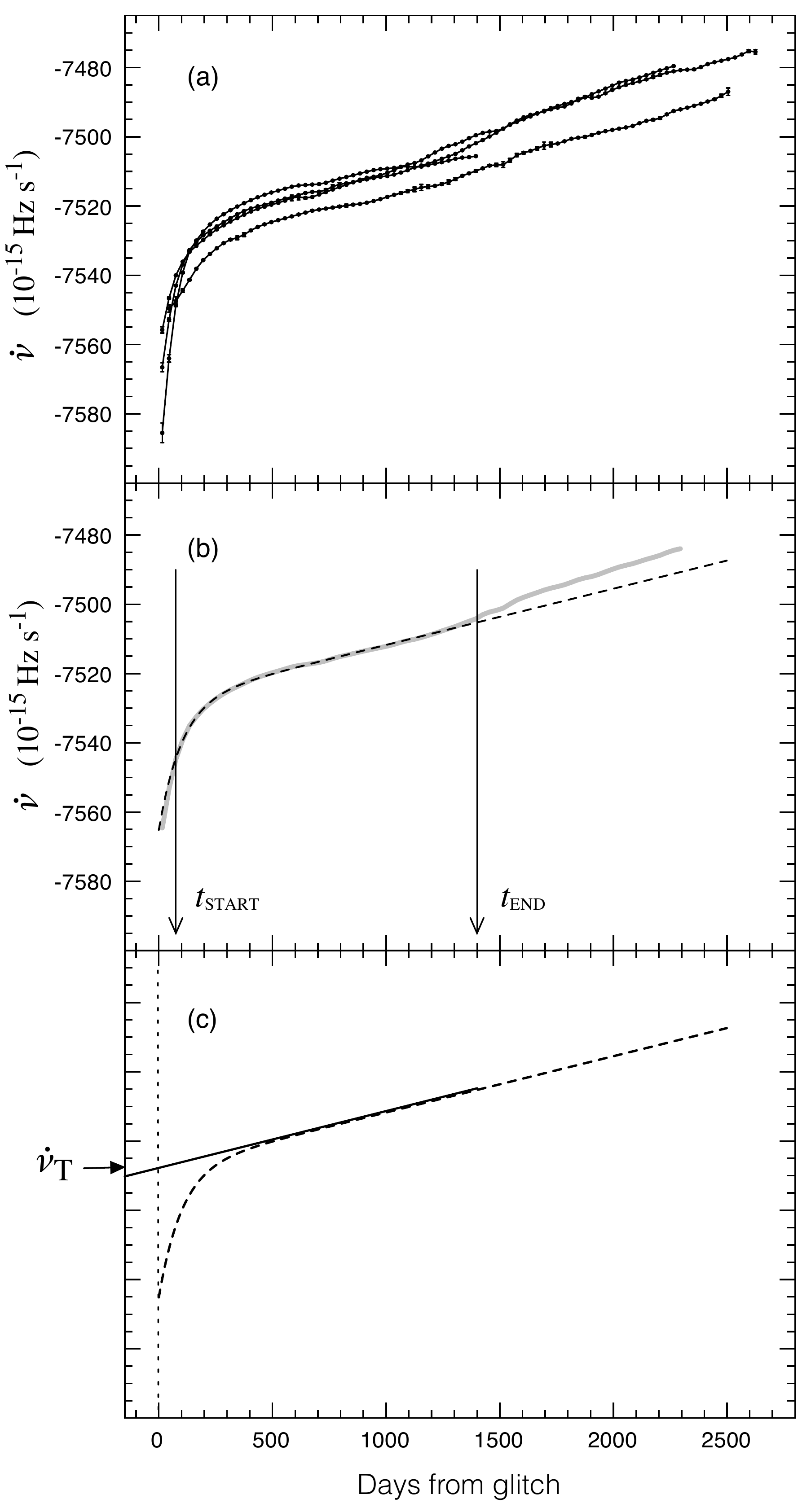}
\caption{Creation of a post-glitch \nudot\ template. 
(a) Superimposed \emph{uniform} sampled 
inter-glitch \nudot\ curves for \psrb.
(b) Glitch \nudot\ template (solid line), obtained by averaging the curves in the top panel. 
The arrows indicate the start and end times of the time-interval over which the template is used to fit the individual inter-glitch curves.
The segmented line is the best fitted function of the form $g(t)=\dot{\nu}_\textsc{T}+\ddot{\nu}_\text{ig}t+\dot{\nu}_\text{d}e^{-t/\tau_\text{d}}$ (Eq. \ref{gmodel}) to the template, between $t_\textsc{start}=75$\,d and end $t_\textsc{end}=1400$\,d.
\cme{(c) Visual representation of $\dot{\nu}_\textsc{T}$, the \emph{mean} value of the template (see text).
The segmented line is $g(t)$ (as in panel (b)) and the solid line is the linear term of that function.
The quantity $\dot{\nu}_\textsc{T}$ is the value (indicated by the arrow) of that line at the glitch epoch. The vertical dotted line marks the glitch epoch.  }
}
\label{tempex}
\end{center} \end{figure}

First, \emph{uniform} inter-glitch \nudot\ datasets, having a homogeneous time sampling, are created via linear interpolation of the original \nudot\ data points (Fig. \ref{tempex}).
Error bars for the \emph{uniform} data points are calculated by propagating the uncertainties of the original \nudot\ measurements.
To superimpose the post-glitch curves each one is shifted in time such that the glitch epoch is at $t=0$.
The first point of a \emph{uniform} dataset is set at a time corresponding to half the  sampling time. 
If there are no data available to perform an interpolation at that time the first point is set one sampling time later.
In the cases of PSRs \bb\ and \cc a sampling interval of 30 days was used, which corresponds to twice the sampling rate of the original \nudot\ data, and the first \emph{uniform} datapoint was set 15 days after each glitch.
In the few cases in which this was not possible the first point was set one sampling time later (i.e. 45 days after the glitch).
For the Vela pulsar, \nudot\ data points were calculated every 2 days, in order to properly sample the high cadence available on the early dates of the original \nudot\ dataset.

Then, a post-glitch \nudot\ template is created by averaging the \emph{uniform} \nudot\ curves created for each inter-glitch interval, using the glitch epochs as time origins to align them (see the example in Fig. \ref{tempex}).
The times $t_\textsc{start}$ and $t_\textsc{end}$, which are defined relative to the glitch epoch, are used to define the range over which the template will be used.
In general, they are chosen such that the template is used only in the time range that is covered by most of the inter-glitch curves.
In the example in Fig. \ref{tempex}, three inter-glitch curves start at day 15 and one at day 45. 
Setting $t_\textsc{start}=45$\,d would satisfy the above criteria.
However, as an additional criterion, it might be desirable to start at the next point, $t_\textsc{start}=75$\,d, to avoid using values of the uniform curves that depend on the first analyses after the glitches (those with epochs on the first TOA after the glitch, see \S \ref{curves}).
The reason is that these first points are in general very different between the various inter-glitch curves, thereby making the definition of the template slightly uncertain.
We tried both approaches and obtained consistent results, nonetheless.
In the example, $t_\textsc{end}=1400$\,d corresponds to the end of the shortest inter-glitch curve, the one after the 4th glitch.
Beyond this time the template is defined by the other three inter-glitch curves only.

For each glitch, the template is scaled by a factor $a$ and shifted in \nudot\ by an amount $s$ to fit the uniform post-glitch \nudot\ curve, via a least square process.
We calculate $a$ and $s$ such that they minimise the weighted sum of squares of the residuals,
\beq
\sum_{i=\textsc{start}}^{\textsc{end}}\omega_i R_i^2  \quad ,
\label{minim}
\eeq
where $i$ counts every data point in the interval, $\omega_i$ is the weight assigned to each term, and the residual
\beq
R_i = \left\{\dot{\nu}_\text{uniform}\right\}_i - \left( a\times \left\{\dot{\nu}_{\text{template}} \right\}_i+s \right).
\eeq

The indices \textsc{start} and \textsc{end} relate to the start and end times, 
$t_\textsc{start}$ and $t_\textsc{end}$.
The same $t_\textsc{start}$ and $t_\textsc{end}$ values are used for all the inter-glitch curves for a given pulsar.
The weights are defined as $\omega_i=1/\sigma_i^2$, with $\sigma_i$ the error-bar of the corresponding \nudot\ value in the uniform dataset.

Having found $a$ and $s$ for each glitch, the long-term \nuddot\ is obtained from a linear fit to the points $(t_g,\dot{\nu}_g)$, where the $t_g$ are the times of the glitches and each $\dot{\nu}_g$ is the representative \nudot-value for the respective post-glitch curve, determined by the results of the least square fit:
\beq
\dot{\nu}_g= a\cdot \dot{\nu}_\textsc{T} +s \quad ,
\eeq
with $\dot{\nu}_\textsc{T}$ the \emph{mean} spin-down rate of the template.
We calculate $\dot{\nu}_\textsc{T}$ by measuring the asymptotic slope of the template and evaluating at $t=0$ (see Fig. \ref{tempex}).
This is achieved by fitting the four-parameter function 
\beq
g(t)=\dot{\nu}_\textsc{T}+\ddot{\nu}_\text{ig}t+\dot{\nu}_\text{d}e^{-t/\tau_\text{d}}
\label{gmodel}
\eeq
to the \nudot\ template.
Here, $\ddot{\nu}_\text{ig}$ is the inter-glitch asymptotic slope of the post-glitch \nudot\ curve (used to calculate $n_\text{ig}$) and both $\dot{\nu}_\text{d}$ ($<0$) and $\tau_\text{d}$ are standard single-exponential parameters that characterise the relaxation of the initial spin-down step.

Finally, to calculate the braking index using Eq. \ref{eq:powerLaw}, we also need representative values for the frequency and its first derivative, which we call $\langle\nu\rangle$ and $\langle\dot{\nu}\rangle$. 
To define these quantities we consider the fits performed to build the \nudot\ datasets (\S \ref{curves}). 
For $\langle\nu\rangle$ we use the frequency value given by the fit to the group of TOAs which is the closest to the centre of the total observation span. 
To calculate $\langle\dot{\nu}\rangle$, we perform a linear fit to all the frequency values obtained during the fits and use that slope. 
We note that, because of the positive frequency steps at the glitches, this value is smaller, in absolute terms, than all the \nudot\ values measured through fits to groups of TOAs. 
The values obtained are in Table \ref{results}, where the epochs associated with $\langle\nu\rangle$ are also quoted.

The uncertainty on the braking index will mostly depend upon the scatter of the $(t_g,\dot{\nu}_g)$ points around the fitted straight line, hence 
we use the propagated standard errors of the linear fit to calculate the braking index error-bar. 
Other sources of error, such as timing noise and the uncertainty of the glitch epochs or the \nudot\ values are less significant.
We comment on the uncertainties of the measurements in section \ref{sec:disc1}.

\begin{table*}
\begin{minipage}{120mm} 
\caption{Results of the fit of Eq. \ref{gmodel} to the \nudot\ templates, weighted RMS of the residuals and mean inter-glitch braking indices, $n_\text{ig}$.}
\label{interGs}
\begin{tabular}{lcccccc}
\hline
\multicolumn{1}{c}{Pulsar} 
& \multicolumn{1}{c}{$\dot{\nu}_\textsc{T}$} 
& \multicolumn{1}{c}{$\ddot{\nu}_\text{ig}$}
& \multicolumn{1}{c}{$\dot{\nu}_\text{d}$} 
& \multicolumn{1}{c}{$\tau_\text{d}$} 
& \multicolumn{1}{c}{RMS$_w$} 
& \multicolumn{1}{c}{$n_\text{ig}$}   
\\ 
& \multicolumn{1}{c}{($10^{-15}$\,Hz\,s$^{-1}$)}
& \multicolumn{1}{c}{($10^{-24}$\,Hz\,s$^{-2}$)} 
& \multicolumn{1}{c}{($10^{-15}$\,Hz\,s$^{-1}$)} 
& \multicolumn{1}{c}{(Days)} 
& \multicolumn{1}{c}{($10^{-15}$\,Hz\,s$^{-1}$)} 
& \\
\hline
B0833$-$45	&	-15660.5(3) &	875(5)  & -44.4(3)  &	102(2)   &	0.2	&  	41.5(3)   \\
B1800$-$21	&	-7528.0(2)  &	188(3)  & -37(1)    &	101(4)   &	0.1	&	25.9(4)   \\
B1823$-$13	&	-7313.6(3)  &	155(2)  & -31(8)    &	120(18)  &	0.4	&	29.5(4)   \\
\hline
\end{tabular} \\
{\sc Note.---} Uncertainties (1-sigma) on the last quoted digit are shown between parentheses. 
\end{minipage}
\end{table*}

\section{Results}
\label{sct:results}
The templates which were adjusted to the inter-glitch \nudot\ curves and the $(t_g,\dot{\nu}_g)$ points obtained for each pulsar are shown in Figs. \ref{0833} and \ref{1800}.
The results of the fits of the exponential model in Eq. \ref{gmodel} to the templates are given in Table \ref{interGs}.
Using the slopes $\ddot{\nu}_\text{ig}$ from these fits we calculate mean inter-glitch braking indices ($n_\text{ig}$), also included in this table. 
The mean frequency, mean frequency derivative, measured \nuddot\ and long-term braking index for each pulsar are in Table \ref{results}.
This table also contains the weighted root mean squares (RMS$_w$) of the \nudot\ residuals relative to the templates and the linear correlation coefficients ($r$) of the linear fits to the $(t_g,\dot{\nu}_g)$ points.
Particulars for each pulsar are discussed below.


\subsection{The Vela pulsar (PSR B0833$-$45)}
\label{sct:vela}

\begin{figure*} \begin{center}
\includegraphics[width=17.5cm]{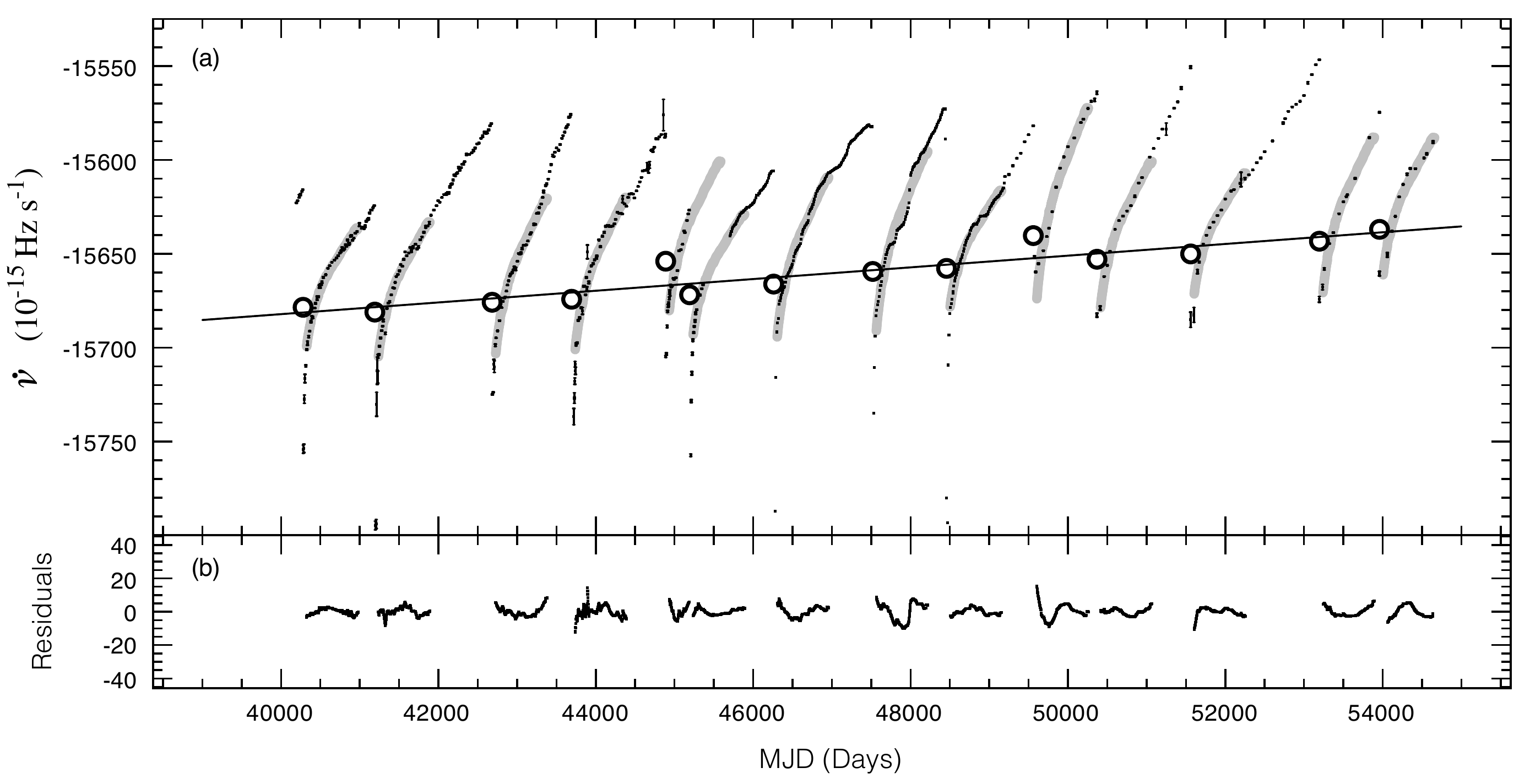}
\caption{The Vela pulsar: glitch \nudot\ template fitting and fit for \nuddot.
(a) The fitted templates (thick grey lines) for each post-glitch curve  superimposed on the \nudot\ time-evolution (black dots). The $(t_g,\dot{\nu}_g)$ points and the straight line that best fits them are also shown, with open circles and a black line.
(b) The residuals $R_i$ for each fit using the same vertical scale as the top panel and the same units.}
\label{0833}
\end{center} \end{figure*}

There are 14 large glitches and their recovery curves included in the dataset, thereby offering good statistics for the linear fit to the $(t_g,\dot{\nu}_g)$ points. 
The average length of the inter-glitch intervals is about $1000$\,d but the shortest interval is only 300\,d long.
While for the earliest glitches the \nudot\ data start a few days after the glitches, for the latest glitches the available data points start about a month after.
The use of a short $t_\textsc{start}$ value could introduce an asymmetry among the $(t_g,\dot{\nu}_g)$ points caused by effects of the short term recoveries of the early glitches, not available for later ones.
Hence we use a template starting at $t_\textsc{start}=45$\,d, which already incorporates in its first point all the available inter-glitch curves.

Because some error bars are missing for the earliest data,
we use the same weights for all values during the template fits (i.e. $\omega_i=1$ in Eq \ref{minim}).
By setting $t_\textsc{end}=700$\,d we obtain a braking index $n=1.7\pm0.2$ and a linear correlation coefficient for the linear fit of $0.920$.
The choice of $t_\textsc{end}=700$\,d ensures the use of a good extent of most of the recovery curves and at the same time avoids the use of the latest parts of the recoveries, which are not present for all curves because of the following  glitch.
The results are plotted in Fig \ref{0833}.

We tested the robustness of the measurement to the values of $t_\textsc{start}$ and $t_\textsc{end}$. 
With $t_\textsc{start}=5$\,d or $15$\,d and $t_\textsc{end}=300$, $1000$ or $1600$\,d, we obtain consistent values, considering uncertainties; with similar (but slightly larger) error bars.
Although it is a mild effect, we note that the best fits (smaller error bars and larger linear correlation coefficients) are those corresponding to $t_\textsc{start}$ and $t_\textsc{end}$ that maximise the level of overlap between all curves.

%

\subsection{\psrb} 
Four large glitches and their recoveries are available for this pulsar in the JBO dataset.
The last glitch occurred in August 2011 and at the moment $\sim1,400$ days of the post-glitch curve are available (compared to more than 2,300 days between the other glitches).
We think, however, that the curve has already acquired some of the characteristic patterns necessary to match the template and decided to include it in the analysis.
The inclusion of a fourth $(t_g,\dot{\nu}_g)$ point in the linear fit should offer a more robust estimate of the braking index.

For $t_\textsc{start}=75$\,d and $t_\textsc{end}=1400$\,d the method produces a linear fit to the $(t_g,\dot{\nu}_g)$ points with linear correlation coefficient  $r=0.948$ and gives a braking index $1.9\pm0.5$.
The fitted templates and $(t_g,\dot{\nu}_g)$ points are plotted in Fig \ref{1800}.
We also tried $t_\textsc{start}=45$\,d and $t_\textsc{end}=1000$ or $2300$\,d and obtained consistent results. 
We adopt the value quoted above (and in Table \ref{results}) because it involves the use of the four inter-glitch curves in equal proportions. 

If only three post-glitch sections are considered ---and trying different values for $t_\textsc{start}$ and $t_\textsc{end}$, the resulting braking index becomes slightly larger (up to $2.0\pm0.9$) but still contained by the error bar of the adopted value.
We note that for this consideration the template was re defined using three glitches.

\subsection{\psrc}
Three full inter-glitch curves are available for this pulsar.
Only a few TOAs are available prior to the first glitch and, although the glitch is clearly detected, it is not possible to obtain reliable measurements of \nudot\ before the glitch.
The best estimate for the epoch of this glitch is indicated with an arrow in
Fig.~\ref{allf1}.
We note that there are two other very small glitches reported for this pulsar \citep{elsk11} that we do not consider here because of the negligible effect they impose on the \nudot\ evolution (they are invisible in Fig. \ref{allf1}).

\begin{figure} \begin{center}
     \includegraphics[width=8.55cm]{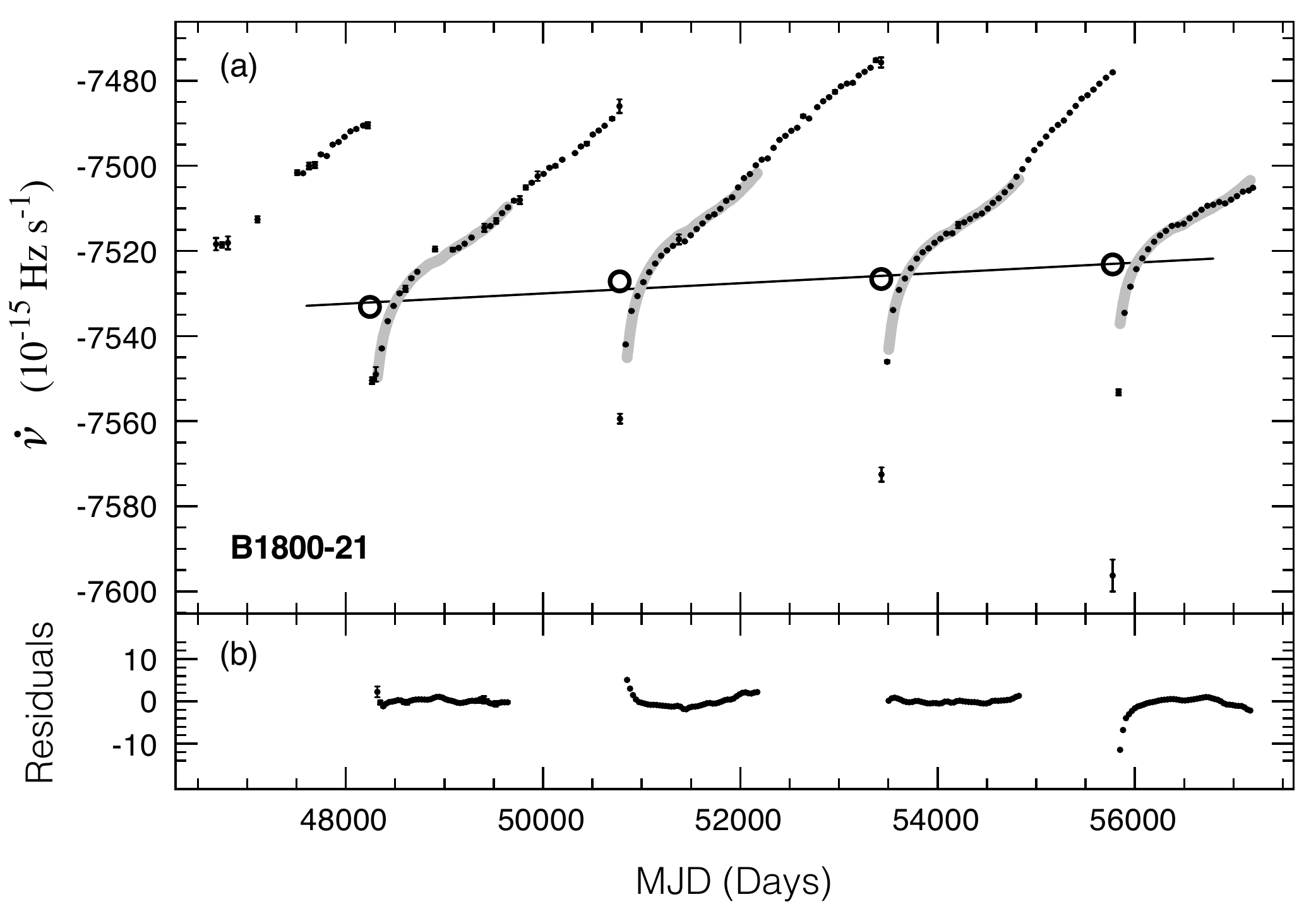}
     \includegraphics[width=8.55cm]{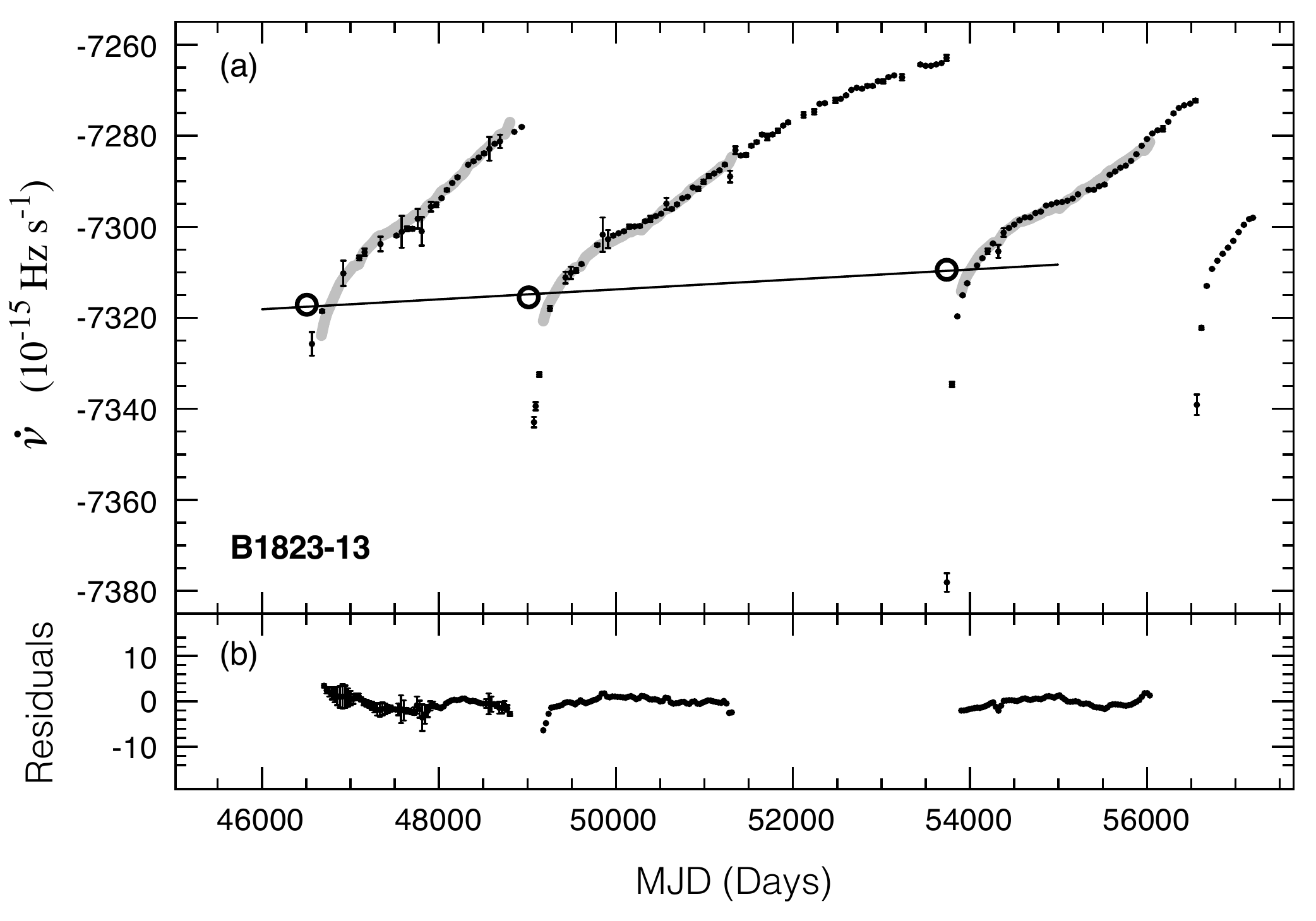}
     \caption{Glitch \nudot\ template fitting and fit for \nuddot. Top: \psrb; bottom: \psrc.
     As in Fig. \ref{0833}.}
     \label{1800}
\end{center} \end{figure}

The first TOAs after the first two glitches are on days 58 and 79, respectively, 
impeding the use of short $t_\textsc{start}$ values.
We choose $t_\textsc{start}=165$\,d and $t_\textsc{end}=2300$\,d, which define the longest possible interval that can involve all three inter-glitch curves completely.
The fitted templates and $(t_g, \dot{\nu}_g)$ points are shown in Fig. \ref{1800}.
This gives a braking index $2.2\pm0.6$, from a linear fit with $r=0.969$.
We also tried all combinations between $t_\textsc{start}=75$ or $105$\,d and $t_\textsc{end}=1000$, $1500$ or $2000$\,d obtaining similar results, with braking indices between $1.4$ and $2.5$, with error bars between $0.5$ and $2.0$.

If the (unfinished) recovery from the fourth glitch is included in the analysis, using the same set of $t_\textsc{start}$ and $t_\textsc{end}$ values, the braking indices are slightly larger in some cases but entirely consistent with the above results.
Furthermore, the braking index for the preferred $t_\textsc{start}=165$\,d and $t_\textsc{end}=2300$\,d is exactly the same, with a smaller error bar (of $0.2$) and a slightly better linear correlation coefficient (of $0.988$).
Nevertheless, we keep the results from the three complete recoveries.

\begin{table*}
\begin{minipage}{127mm} 
\caption{Mean frequencies and frequency derivatives, RMS$_w$ values from the template fits, linear correlation coefficients ($r$) from the fits to the $(t_g,\dot{\nu}_g)$ points, measured long-term $\ddot{\nu}$  values and their associated long-term braking indices.}
\label{results}
\begin{tabular}{llcccccc}
\hline
\multicolumn{1}{c}{Pulsar} 
& \multicolumn{1}{c}{Epoch}
& \multicolumn{1}{c}{\textlangle$\nu$\textrangle} 
& \multicolumn{1}{c}{\textlangle $\dot{\nu}$\textrangle} 
& \multicolumn{1}{c}{RMS$_w$} 
& \multicolumn{1}{c}{$r$} 
& \multicolumn{1}{c}{$\ddot{\nu}$} 
& \multicolumn{1}{c}{$n$}   
\\ 
& \multicolumn{1}{c}{(MJD)} 
& \multicolumn{1}{c}{(Hz)} 
& \multicolumn{1}{c}{($10^{-15}$\,Hz\,s$^{-1}$)} 
& \multicolumn{1}{c}{($10^{-15}$\,Hz\,s$^{-1}$)} 
& 
& \multicolumn{1}{c}{($10^{-24}$\,Hz\,s$^{-2}$)} 
& \\
\hline
B0833$-$45	& 47419 & 11.200  &	-15375(1) &	3.1	&	0.920	&	36(4)  &	1.7	$\pm$	0.2	\\
B1800$-$21	& 51917 & 7.4825  &	-7371(3)  &	0.8	&	0.948	&	14(3)  &	1.9	$\pm$	0.5	\\
B1823$-$13	& 51890 & 9.8549  &	-7191(3)  &	0.8	&	0.969	&	11(3)  &	2.2	$\pm$	0.6	\\
\hline
\end{tabular} \\
{\sc Note.---} Uncertainties (1-sigma) on the last quoted digit are shown between parentheses. 
\end{minipage} 
\end{table*}

\section{The braking index of other glitching pulsars}
\label{others}
In this section we address the braking indices of six other pulsars in Fig. \ref{allf1}, for which we believe our method can give reliable insight into their long-term spin evolution.
Relative to PSRs \bb\ and \cc, these pulsars present somewhat disadvantageous conditions for the use of glitch \nudot\ templates.
In general, they have noisier \nudot\ evolution, irregular post-glitch curves and, for some of them, less than three full glitch recoveries are available.

Initially, the method was applied to all the pulsars in Fig. \ref{allf1} that have undergone at least three medium-to-large glitches (in terms of both $\Delta\nu$ and $\Delta\dot{\nu}$) in the observed periods.
The only pulsars not analysed were PSRs B2334+61 and J1809$-$1917, because both have only one recorded glitch. 
They are, however, fine examples of pulsars for which the method will successfully work, provided there are observing programs which continuously monitor their rotations during the next decades.

The \nudot\ curves of most of the other pulsars are characterised by quasi-random oscillations caused by timing noise, which partially obscure the underlying behaviour.
The irregularity of the variations makes the definition of a template more complicated and undermines the effectiveness of the least squares process.
This produces \nudot\ residuals which are 2-3 times larger (as a fraction of the \nudot\ range covered by the rotation of each pulsar) than those of Vela and the other two pulsars analysed in the last section.

Furthermore, their glitch recoveries are different from those exhibited by Vela and the other two pulsars.
With the exception of PSR B1727$-$33, the fits of Eq. \ref{gmodel} to the \nudot\ glitch templates of these pulsars are unconstrained because the exponential parts of their recoveries are either masked by the noise or have a different shape (like PSR J2229+6114).
Therefore, to define $\dot{\nu}_\text{T}$ we performed linear fits to the templates and used the extrapolated value of the fitted line at the glitch epoch.
This keeps the spirit of the previous definition, as it also corresponds to the asymptotic trend of the recovery extrapolated back to the glitch epoch.

Below we describe the results obtained for the six aforementioned pulsars.
It was possible to obtain, for the first time, long-term braking index estimates for four of them. 
In all cases the choices of $t_\textsc{start}$ and $t_\textsc{end}$ were such that they completely covered the shortest of the post-glitch curves, so all curves are used equally.

\begin{table}
 \caption{Other long-term braking index estimates determined in this paper.}
 \label{tbl:enes2}
 \begin{tabular}{lll}
  \hline
  \multicolumn{1}{l}{Pulsar} & \multicolumn{1}{l}{J name} & \multicolumn{1}{c}{$n$}  \\
  \hline
   B1727$-$33  & J1730$-$3350   & $1.8 \pm 0.3$    \\
   B1737$-$30  & J1740$-$3015   & $1  \pm  1 $    \\
   B1757$-$24  & J1801$-$2451   & $1.1 \pm 0.4$    \\
   J2229+6114 & J2229+6114   & $0  \pm  1 $    \\
  \hline 
 \end{tabular} \\
\end{table}

PSR J0631$+$1036 has glitched 15 times since the start of the observations and most of the events are small, with
13 of them having sizes $\Delta\nu\leq 0.2\,\mu$Hz.
The two largest glitches ($\Delta\nu\sim6$, $11\,\mu$Hz) are not as large as the Vela glitches but involve relatively large \nudot\ steps (see Fig \ref{allf1}).
Significant \nudot\ steps are also induced by some of the other medium sized glitches.
This is rare and these steps could be caused by timing noise, rather than glitches.
If we use the five events with the largest \nudot\ steps we obtain $n=8\pm5$.
Alternatively, if we only use the three events with the largest \nudot\ steps we obtain $n=4\pm1$.
In both cases the residuals are rather large, accounting for about 30\% of the dispersion of the \nudot\ evolution.

There are three large glitches detected for PSR B1727$-$33 but the recovery from the third glitch is not complete yet.
This is a Vela-like pulsar in many respects and a reliable braking index measurement will soon be possible.
Despite the noise, the recoveries do show some exponential behaviour and it was possible to fit the model in Eq. \ref{gmodel} to define $\dot{\nu}_\text{T}$. 
The first glitch occurred just before the start of the JBO observations (Fig. \ref{allf1}) \citep{jml+95} but its recovery is well sampled since the beginning.
We obtained a good linear fit, indicating a braking index $n=1.8\pm0.3$, but rather large \nudot\ residuals.

PSR J1737$-$30 is one of the most active glitching pulsars known \citep{ml90,zwm+08,elsk11}, with 33 glitches recorded in $\sim30$ years.
Its glitch size distribution is very different to those of the Vela-like pulsars, exhibiting a large number of small and medium size glitches and a rather low probability for large glitches \citep{elsk11}. 
The largest glitch ($\Delta\nu=4.4\,\mu$Hz; MJD 55213) is smaller than most of the Vela glitches and its \nudot\ evolution is rather flat, with no more than  0.4 \% variation in 30 years. 
This already suggests a very low $\ddot{\nu}$.
We apply the method over the 9 largest glitches and obtained $n=1\pm1
$.
The use of more or less glitches offers similar results, all of them indicative \cme{(with considerable uncertainties)} of a braking index between 0 and 1.

In about $24$~yr, \psra\ has experienced four large glitches, a frequency similar to Vela's (every 3-4 yr), and it also exhibits similar \nudot\ patterns, though highly contaminated by timing noise (Fig.~\ref{allf1}).
Despite this, the \nudot\ residuals resulting from the template fitting are not too large and the linear fit is relatively good, giving a braking index $n=1.4\pm0.4$.

The results for PSR J2021+3651 are unclear, probably because it has not been observed for long enough and also because the recoveries are very different from glitch to glitch. 
The template fitting gave the largest residuals of all the pulsars analysed and small changes of the $t_\textsc{start}$ and $t_\textsc{end}$ parameters gave different results, with braking indices ranging from $7$ to $10$.

PSR J2229+6114 has seven detected glitches, with a variety of sizes.
The recoveries are similar but not exactly repetitive and do not resemble those of Vela and the other pulsars (Fig. \ref{allf1}).
We defined a template using the post-glitch curves of the two largest glitches and used the three largest glitches to measure the braking index.
The template fitting produces very small \nudot\ residuals and gives $n=0\pm1$.

\section{Discussion}
The braking index we obtain for the Vela pulsar is only slightly larger than the measurement of $n=1.4\pm0.2$ reported by \citet{lpgc96}, which was obtained by considering data between 1969 and 1994 (9 glitches), and entirely consistent with the value $n=1.6\pm0.1$  reported by \citet{dlm07}, who used a similar method to the one used by \citet{lpgc96} for data between 1981 and 2005  (10 glitches).

The other two pulsars for which solid measurements were possible were chosen because three or more large and regular glitches and their full recoveries are included in the available data, where this is a minimum value that comes from the requirement to have at least three $(t_g,\dot{\nu}_g)$ points to perform a linear fit to obtain $\ddot{\nu}$.
How certain can one be that the use of only three or four points (three or four glitches) will give relevant information about the long term \nudot\ evolution of a given pulsar? What happens if a pulsar shows deviations from the regularity exhibited by Vela and the other two pulsars?
In \S~\ref{sec:disc1} we analyse the behaviour of the Vela pulsar trying to shed light on these issues.

The new measurements of long-term braking indices are all similar and are consistent with values of $n\sim2$.
In \S \ref{sec:disc2} we discuss the implications of this result, while in \S \ref{sec:nig} we comment on the short term behaviour between glitches. 
A discussion about the role that glitches and superfluid dynamics might have on the observed short and long term behaviour is presented in \S \ref{sec:disc3}.

%

\subsection{The method}
\label{sec:disc1}
With 14 available glitches, the braking index measurement for the Vela pulsar is very robust. 
Indeed, the measured value is almost insensitive to variations of the few parameters involved in the method (\S \ref{sct:vela}).
For PSRs \bb\ and \cc, with four and three full recoveries respectively, the results obtained for various values of $t_\textsc{start}$ and $t_\textsc{end}$ are all consistent.
Interestingly, the best fits are obtained when $t_\textsc{start}$ and $t_\textsc{end}$ are chosen such that all inter-glitch curves are equally used, i.e. using templates as long as the shortest post-glitch \nudot\ curve.

To assess the long-term predictive power of the method when using only four glitches we take advantage of the long dataset available for the Vela pulsar and the large number of glitches it contains. 
It should be noted, however, that every pulsar exhibits a different behaviour (in terms of glitch recoveries, glitch timescales and timing noise properties), hence the conclusions presented below can only be used as a qualitative reference to understand the behaviour of other pulsars. 

\begin{figure} \begin{center}
\includegraphics[width=8.0cm]{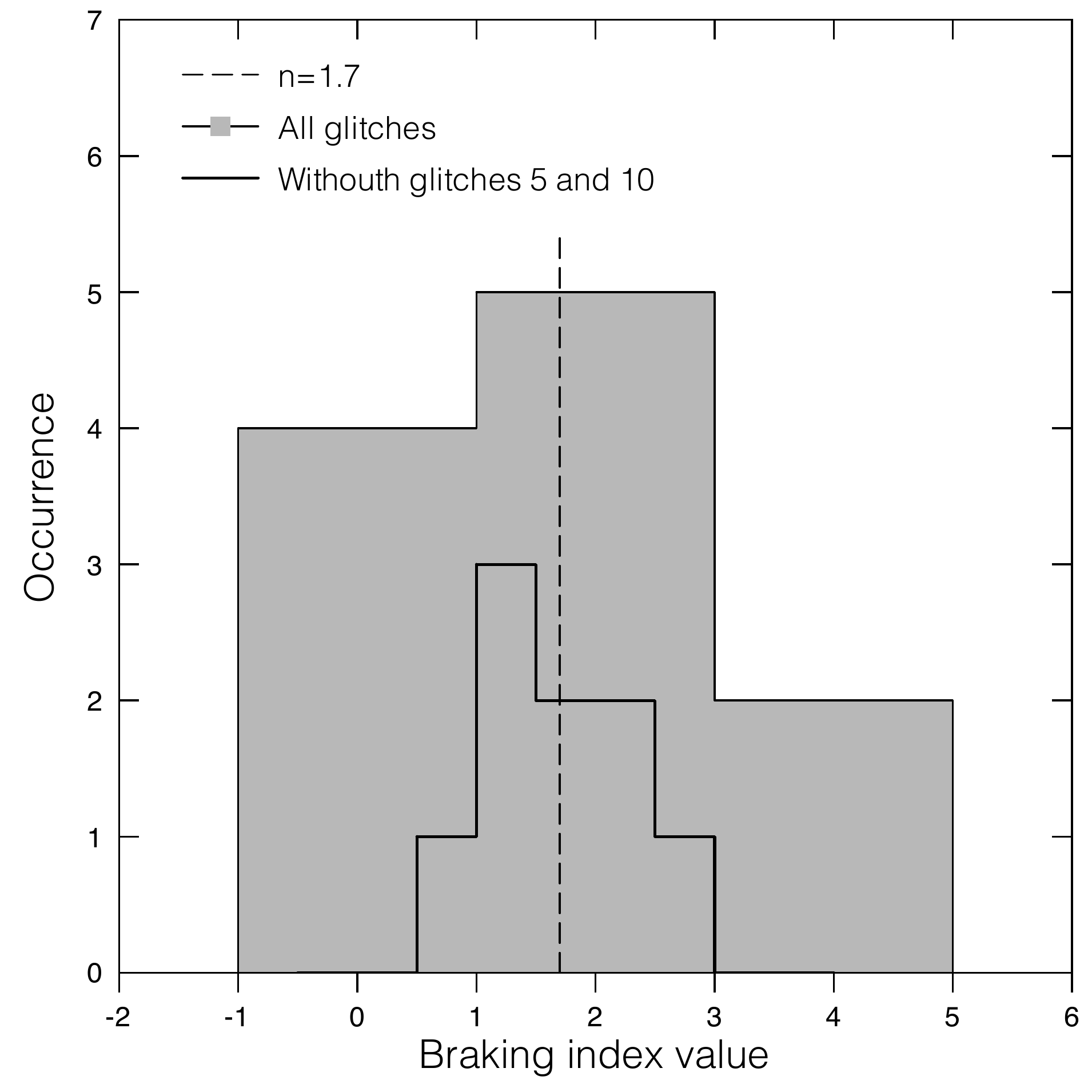}
\caption{Histograms for braking index values calculated using groups of 4 glitches in the Vela pulsar. The grey-filled histogram is for all glitches.
Glitches 5 and 10 were eliminated to generate the histogram with smaller bin sizes.}
\label{histoenes}
\end{center} \end{figure}

We have calculated 11 local braking indices for the Vela pulsar using groups of four contiguous glitches \cme{(note that these values are not independent). }
The measured values distribute around their average of $1.9$ with a standard deviation of $2.3$ (Fig. \ref{histoenes}) and have error bars of size $1.7$ on average.
This indicates that, in this case, the rotational evolution over 4 glitches can deviate from the long-term trend by an amount equivalent to $1$ or up to $2$ braking index units.
For instance, if we were to have data for the Vela pulsar only from before the first glitch and until just before the fifth glitch, we would obtain a braking index of $0.9$. 
Similarly, between glitches 6 and 9 we would measure a braking index of $2.5$.
However, further considerations are necessary in order to understand this better.


There are two $(t_g,\dot{\nu}_g)$ points, corresponding to the recoveries of Vela's glitches 5 and 10, which are considerably offset from the best fit straight line in Fig. \ref{0833}.
Glitch 5 is the smallest, by at least a factor of two, of the large glitches in the Vela pulsar\footnote{The smallest glitch reported has $\Delta\nu\sim0.1\,\mu$Hz \citep{cdk88} which is invisible in the \nudot\ plots, hence not considered in this analysis.} (with $\Delta\nu\sim13\,\mu$Hz) and glitch 10 actually consists of two glitches separated by 32 days only \citep{fla94,fm94,bf11}.
Their combined size ($\Delta\nu\sim12\,\mu$Hz) is similar to glitch 5 and therefore small compared to all other glitches (\S \ref{sec:largeglits}).
Therefore, the two smallest glitches in the dataset (glitches 5 and 10) have recoveries for which their measured $\dot{\nu}_g$ values are smaller (in absolute value) than the prediction of the best linear fit to the $(t_g,\dot{\nu}_g)$ points.
This could happen because their modest sizes involve a different response of the star and the rotation is not reset to the same basal configuration that larger glitches can reach.

If glitches 5 and 10 are left out of the analyses and the 4-glitch partial braking indices are recalculated (for instance one of them would use glitches 3, 4, 6 and 7, skipping glitch 5) the dispersion around the mean decreases and the uncertainty of the single measurements significantly improves.
The average value of these 9 analyses is $1.7$, with a dispersion of $1.8$ and average error bars of $0.3$ (Fig. \ref{histoenes}).
In this case (Vela), the main trend is dominated by the effects of the largest glitches and it appears that one could prescind from the smaller events when using our method to measure the braking index.
In general, if the glitch sizes in a given pulsar show no regularity, and the spin-down evolution consists of a series of unequal glitch responses, one should expect a larger scatter of the $(t_g,\dot{\nu}_g)$ points around the best fit straight line. 

Indeed, in the cases of PSRs J0631$+$1036 and B1737$-$30, which present the two broadest size distributions, the scatter is comparable to the intrinsic \nudot\ variations the pulsars exhibit.
Furthermore, the spin-down steps at the glitches are not very large in these two pulsars, thereby modifying the spin-down evolution only mildly. 
Nonetheless, provided long enough observational coverage is available, the series of $(t_g,\dot{\nu}_g)$ points should still reflect the long-term evolution of the pulsar, as we believe is the case of PSR B1737$-$30 (we used 9 glitches covering $\sim27$\,yr).
For the cases of PSRs \bb\ and \cc, considering that they exhibit superior regularity in their glitch sizes and recoveries, combined with timing noise levels lower than the Vela pulsar, the quoted braking indices and their uncertainties are likely good estimates of their long-term underlying trends.

The uncertainties for the long-term braking indices, reported in Tables \ref{results} and \ref{tbl:enes2}, were propagated from the formal errors of (unweighted) linear fits to the $(t_g,\dot{\nu}_g)$ points.
We believe that the reported error bars reflect well the uncertainty of using this method to determine the underlying slope of \nudot\ from the the currently available data and that they are a good approximation to 
the uncertainties relative to the real long-term behaviour, which will only be revealed after many more glitches (and years of monitoring).
\cme{
Entirely consistent values are obtained if weighted fits are used instead.
The error bars obtained from the weighted fits, after they are multiplied by the reduced $\chi^2$ values, are closely the same as the ones from the unweighted fits.  
The only exceptions are PSR \bb, for which the uncertainty decreases to $0.2$, and PSR B1737$-$30, for which the error bar doubled. 
To define the weights we used the RMS values from the template matching, which should somewhat represent the effects of timing noise in our measurements. 
The results obtained suggest that these effects are not larger than the intrinsic dispersion of the post-glitch curves around a linear underlying trend. 
Considering the very small number of data points participating in most of these fits, and the consequent risk of ending up determining the general slope with two points only, we prefer the unweighted fits for now. 
} 


\subsection{Low braking indices: implication for motion in the \ppdd}
\label{sec:disc2}
\begin{figure*} \begin{center}
\includegraphics[width=15.0cm]{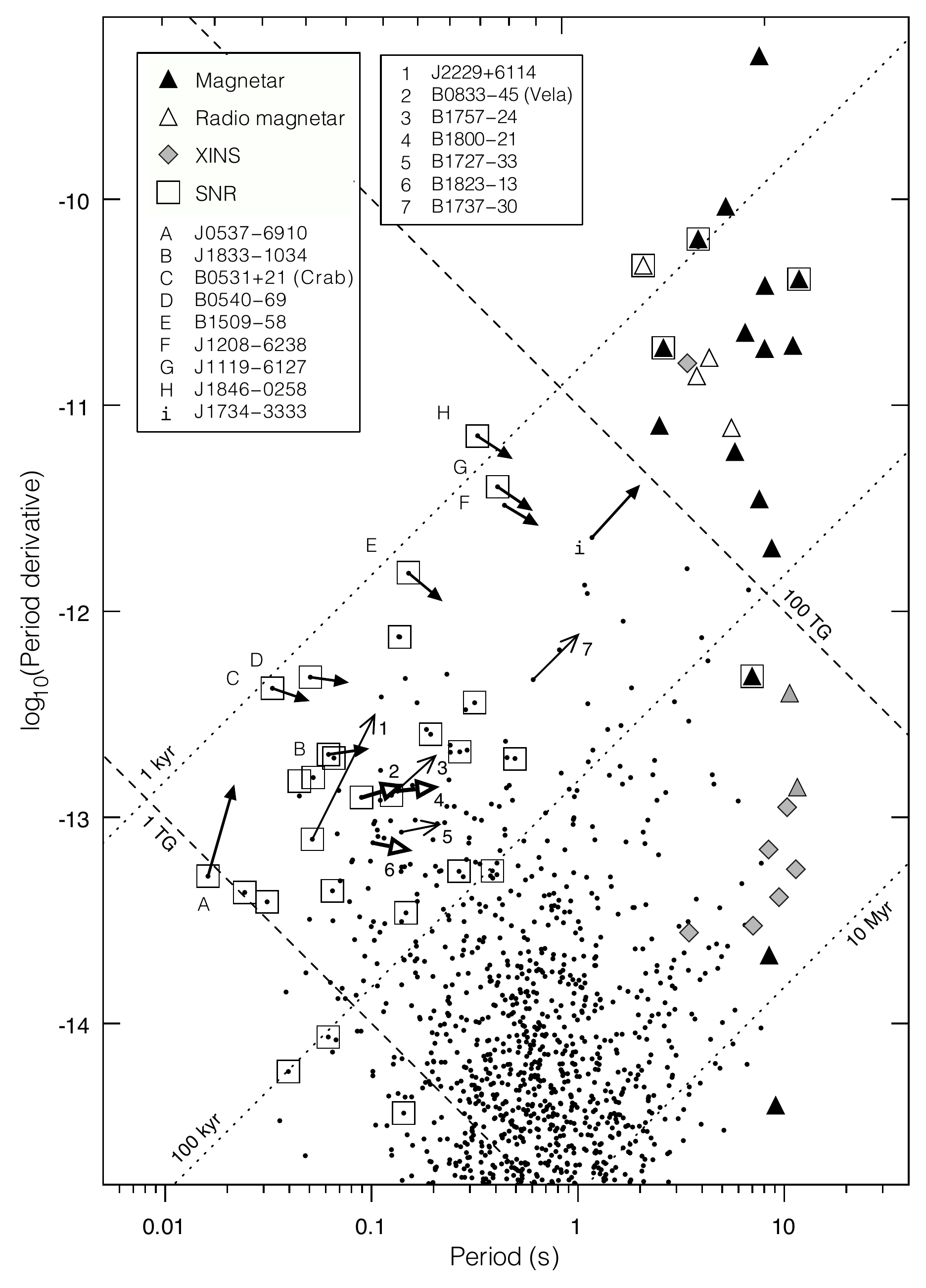}
\caption{The $P$--$\dot{P}$ diagram. 
Equal logarithmic scales are chosen so that a pulsar will move with a slope of $2-n$, as indicated by the arrows. 
Letters (A to I) are used to label pulsars with previously reported braking indices and numbers (1 to 7) are used for the new measurements.
The lengths of the arrows indicate the positions of these pulsars after $1\,\tau_c$ (for a constant $n$; excepting the arrow for PSR J0537$-$6910, which stops at $2$\,kyr because in this case $P$ becomes zero before $1\,\tau_c$. 
Measurements between glitches ($n_\text{ig}$) are used when $n$ is not available (Table \ref{tbl:enes}).
White headed, thick arrows are used for PSRs \bb, \cc\ and Vela.
Slim open arrows are used for the estimates in Table \ref{tbl:enes2}.
Lines of constant magnetic field are dashed and lines of constant characteristic age are dotted.
Information regarding associations and most rotational parameters were taken from the ATNF Pulsar catalogue \citep[\url{http://www.atnf.csiro.au/research/pulsar/psrcat/}; ][]{mhth05} and the McGill Online Magnetar Catalog \citep[\url{http://www.physics.mcgill.ca/pulsar/magnetar/main.html}; ][]{ok14}.
The two grey triangles are magnetars for which only upper limits for their $\dot{P}$ values are known.
}
\label{ppdot}
\end{center} \end{figure*}

One practical way to understand long-term braking indices, independently of the braking mechanisms or processes giving rise to their values, is by using the \ppdd; a plot of $\dot{P}$ versus $P$ for all known pulsars (Fig. \ref{ppdot}).
In this diagram the different populations of pulsars are easily visualised and their possible evolutionary connections can in principle be evaluated, since pulsars move from left to right with a slope given by $2-n$. 
Therefore, braking index measurements represent the long-term actual motion of pulsars across the diagram.

Our measurements indicate that Vela, PSR \bb\ and PSR \cc\ move almost horizontally on this diagram (white headed, thick arrows in Fig. \ref{ppdot}).
This is the mean trend they have followed during the span of the observations 
and it is significantly different to the slope of -1 predicted by the standard $n=3$ evolution.
We have shown that other glitching pulsars might also be evolving with low braking indices (Table \ref{tbl:enes2}) and it is natural to think that maybe most Vela-like pulsars evolve in a similar fashion.
An immediate consequence would be that an important fraction of all pulsars with $\tau_c<100$\,kyr could be moving up or to the right in the \ppdd\ (slopes $2-n\geq0$; see Fig. \ref{ppdot}). 

On the other hand, younger pulsars like the Crab pulsar move with slopes closer to $-1$ according to their measured $n_\text{ig}$ values, which are similar to the long-term $n$ (Table \ref{tbl:enes}; \S \ref{sec:nig}).
The direction of their movement on the diagram somehow matches the general (though perhaps too simple) picture, in which pulsars are born at the top left of the diagram, evolve with braking indices $\sim3$ and after about $1$\,Myr join the main pulsar population (i.e. $P\sim0.5$\,s and $\dot{P}\leq10^{-14}$).
However, if they were to turn into Vela-like pulsars the story may be different. 

While it seems plausible that the Crab and other very young pulsars were indeed born at the top left of the diagram (near their current positions), the place where Vela-like pulsars come from is more uncertain.
Either they are the descendants of pulsars like the Crab pulsar or they were born somewhere else, to the bottom left of where they are now.
%
Is there a time in a pulsar's life at which its glitch activity increases and its braking index decreases? 
Or, alternatively, are there two (or more) classes of pulsars that evolve differently?
Some fundamental differences between neutron stars could in principle produce different glitching properties \citep[like their masses and temperatures,][]{heaa15}, though it is unclear whether this could generate separate populations. 
Different levels of accretion of fall-back material after the supernova explosions could momentarily bury the magnetic fields and also produce different spin evolutions, as the fields re-emerge \citep[e.g.][]{ho2015}.
However, it might also be possible that glitch activity is the main driver of the spin evolution of young pulsars (see \S \ref{sec:disc3}) and that the glitch mechanism in very young pulsars is not fully mature, yet.
Hence, maybe we are looking at different stages of a single story and the Crab pulsar will indeed become a Vela-like pulsar. 
As long as the monitoring of a good number of young pulsars continue, we might be able to answer these questions in the future. 

The future positions of Vela-like pulsars in the \ppdd\ depend strongly on the timescales during which the main process that produces the observed $n$ values can remain effective. 
If the current motion depends mainly on their high glitch activity, then the slopes of their tracks might decrease in the future, because the glitch activity of all pulsars with longer periods (or smaller $|\dot{\nu}|$) is generally low. 
Their tracks might also turn down if the evolution is driven by magnetic field re-emergence or other processes related to magnetic field evolution, as some calculations suggest \citep{gc15,ho2015}.
However, maybe deeply buried or very strong magnetic fields (or both) could keep a pulsar moving up on the diagram for longer times (consider PSR J1734$-$3333, see below).
%
Regardless, it appears that Vela-like pulsars could join the bulk of the main population only if 
their tracks were about to change and the slopes of their motions rapidly became very negative (i.e.  if  their long-term braking indices became much larger than three). 
Otherwise, their final positions on the diagram are more likely to be located to the right and top of the main population. 


%

It seems worth to wonder, therefore, for the true origin of the pulsars which now form the main population.
As discussed above, perhaps very young pulsars (like the Crab pulsar) are indeed different to Vela-like pulsars and will continue to evolve with somewhat negative slopes, eventually joining the bulk of the main pulsar population. 
Another group of young pulsars which could give origin to at least part of the main population are the Central Compact Objects (CCOs; young radio quiet neutron stars at the centre of SNRs), which reside almost at the bottom and to the left of the main cloud of pulsars \citep[not visible in Fig. \ref{ppdot}. See][]{ho11,gha13,ghak13}.







\subsubsection{Characteristic ages and the case of PSR J1757$-$24}
Another consequence of having a population of pulsars moving with slopes $\geq0$ across the \ppdd\ is that their characteristic ages ($\tau_c$) change at rates slower than $1$\,yr per yr, 
hence they could largely underestimate the true ages.
Assuming that their initial spin periods are significantly shorter than the current ones, we calculate ages $\tau\simeq30$-$35$\,kyr (Eq. \ref{eq:age}) for the Vela pulsar and PSRs \bb\ and \cc.
Instead, if the initial periods are taken as $50\%$ of the current values the ages become smaller ($10$-$20$\,kyr).\footnote{Long initial periods work on the opposite direction to low braking indices and tend to make $\tau_c$ overestimate the true ages.}
In general, low braking indices can help to accommodate the discrepancies observed between $\tau_c$  and the age of the SNRs in various systems \citep{gf00,ka14}.
In particular, for the Vela pulsar this result gives preference to those calculations indicating a SNR's age much older than $10$\,kyr (\S \ref{sct:velas}). 

The case of PSR \aaa\ is particularly interesting.
This pulsar powers the radio/X-ray PWN G5.27$-$0.9 and could be associated with the SNR G5.4$-$1.2 \citep{ckk+87}.
The whole system is known as "the Duck" due to the morphology of the gas distribution and might constitute the spectacular view of a pulsar making its way out of the SNR into the interstellar medium \citep[see][]{bgc+06}.
However, taking into account the rather small upper limit for the pulsar's proper motion and some morphology considerations, \citet{gf00} concluded that the association would be possible only if the age of the system was significantly older than $\tau_c=15.5$\,kyr.
They indicate that this would be the case if PSR \aaa\ was evolving with a braking index $n<1.3$, which nicely coincides with our detection of $n=1.1\pm0.4$ (Table \ref{tbl:enes2}). 
\citet{bgc+06} used new observations to constrain the proper motion and updated this result indicating that the age of the system must be older than $70$\,kyr.
Assuming an initial spin period of $10$\,ms\footnote{The current period of PSR \aaa\ is $125$\,ms.} and using the measured braking index it is possible to obtain an age of $70^{+50}_{-25}$\,kyr, where the uncertainties correspond to the maximum and minimum $n$ values allowed by its error bar.
But this might be a rather short initial period, even when compared to what has been concluded for short period pulsars like the Crab \citep[$\sim20$\,ms;][]{fk06,ljg+15}.
We note, however, that it is unlikely that the pulsar has evolved with a constant braking index since its formation and that these calculations are a simplification of the real evolution of the pulsar.
While a previous evolution with $n\sim3$, from birth until today, would add very little time ($\sim10$\,kyr) to the above age estimate (for reasonable initial spin periods), if the pulsar were born with a smaller $\dot{P}$ and evolved with an even lower braking index (as PSR J0537$-$6910; Table \ref{tbl:enes}, Fig. \ref{ppdot}), then it could be much older and it would be possible to satisfy the age requirement to ensure the association with the SNR.
For example, a pulsar born with a period of $80$\,ms and $\dot{P} \sim10^{-14}$ would reach the period of PSR \aaa\ in $100$\,kyr, if it evolved with $n\sim0.5$.
Such an evolution might not be rare and it could be driven by magnetic field evolution or long-term dynamical superfluid effects, as we describe below \citep[e.g.][]{ha12,gc15,ho2015}. 
We conclude that, at least in terms of age and spin evolution, the association of the pulsar with the SNR is feasible.

\subsection{The spin evolution between glitches}
\label{sec:nig}
The Vela pulsar and PSRs \bb\ and \cc\ 
evolve with very large inter-glitch frequency second derivatives, which produce high inter-glitch braking indices $n_\text{ig}>25$ (Table \ref{interGs}).
These short-term trends are superimposed to the long-term evolution, which is described by $n\sim2$.

This is a generic behaviour of Vela-like pulsars. 
Indeed, all the pulsars in Fig. \ref{allf1} that show Vela-like glitch recoveries exhibit large $n_\text{ig}$ values (the smallest is $n_\text{ig}\sim 10$, for PSR B2334+61). 
The cases of PSRs J0631$-$1036 and B1737$-$30 seem different because of their less regular behaviour.
Nonetheless, in the time intervals in which it is possible to measure \nuddot, we find $n_\text{ig}$ values which are also greater than three.
The large $n_\text{ig}$ values are likely a byproduct of the large glitches and might represent a slow term of the recovery process. 

All younger pulsars known, with $\tau_c\sim1$\,kyr\footnote{PSRs B0531$+$21, B0540$-$69, J1119$-$6127, J1208$-$6238, B1509$-$58 and J1846$-$0258.}, exhibit inter-glitch braking indices which are similar to their long-term trends (Table \ref{tbl:enes}). 
In general, they also exhibit smaller glitches than Vela-like pulsars. 
The only two large glitches ever detected in the monitored rotation of these pulsars (for PSR J1119$-$6127 and J1846$-$0258) showed peculiar properties \citep[see \S \ref{intro};][]{awe+15,akb+15} and, although the influence they might have imprinted on the long-term evolution is still unclear, their recoveries have already evolved to \nuddot\ values which imply $n_\text{ig}<3$. 
While these two pulsars exhibit standard glitch activities for pulsars of similar \nudot\ (entirely due to the presence of the two large glitches), the glitch activities of the Crab pulsar and the other very young pulsars are significantly lower than that of Vela-like pulsars \citep{elsk11}. 
Therefore, the relatively low $n_\text{ig}$ values of very young pulsars
could be the result of a braking mechanism which dominates over the effects of a somewhat different (perhaps younger) glitch activity.

\subsection{Understanding low braking indices in terms of superfluid dynamics}
\label{sec:disc3}
Timing observations tell us about the rotation of the crust of the neutron star, which slows down due to magnetic or wind braking, or both. 
While a very energetic wind could reduce the braking index to some extent \citep[e.g.][]{mel97,hck99,txsq13}, the braking index could also be reduced by the emergence of a buried stronger magnetic field (the latter alternative  is briefly discussed by the end of this section). 
Glitches represent some superfluid component which gives angular momentum to the crust in the form of bursts. 
The \nudot\ steps at glitches and the recoveries are the superfluid response to the glitches and, over time, they could reduce \nuddot\ and $n$ too. 

Thus identifying the main process that causes an observed braking index can be a problem with more than one possible solution.
With only timing observations, in general we would be unable to discern between two or more braking mechanisms.
Nonetheless, the observation of intermittent pulsars and mode changing pulsars with measured $\dot{P}$ changes gives us, at least, means of exploring the contributions of the magnetic and wind braking mechanisms \citep{klo+06,lhk+10}. 
For our sources, one can only consider those options which seem more likely to affect their rotation. 

The paths of the Vela-like pulsars on the \ppdd\ are dramatically modified by large and negative $\Delta\dot{\nu}$ steps at the glitches, thereby producing effective long-term motions with slopes just above zero. 
Therefore, it is possible that the long-term braking indices measured for these pulsars are not representative of whichever braking mechanism is in action but of the cumulative effects of their glitch activity.

We now briefly explore the hypothesis that low braking indices in Vela-like pulsars are the secondary effect of their large glitch activity.
This idea is based on the fact that a series of large and negative $\Delta\dot{\nu}$ steps, which are not fully reversed by the recoveries, might involve a systematic decrease of the effective moment of inertia over which the braking torques are applied. 
This would consequently reduce the observed braking index, as can be concluded from the following relationship:

\begin{equation}
\label{idot}
\frac{\dot{I}}{I}=\frac{n_\text{obs}-n_0}{2\tau_c} \quad ,
\end{equation}
where $\dot{I}$ is the time derivative of the moment of inertia $I$ (of all matter corotating with the crust), $n_\text{obs}$ is the observed braking index and $n_0=3$, assuming an underlying magnetic braking.
This expression comes from differentiation of Eq. \ref{eq:powerLaw} while allowing $\kappa$ to vary with time and assuming pure magnetic braking by a constant magnetic dipole \citep[e.g.][]{ls90}.
Thus a negative $\dot{I}$ implies $n_\text{obs}<3$, as observed.

In most models, glitches are produced by a neutron superfluid component inside the star which cannot follow the spin-down of the crust and acts as an angular momentum reservoir [we base our arguments on a two fluids model but other variants \citep[e.g.][]{aaps84,wm08} should also be considered; see \citet{hm15} for a review of glitch models].
In general, the angular momentum of a superfluid is quantised and stored in vortices of equal circulation.
Because the total angular momentum is proportional to the density of vortices, the superfluid will spin down by generating a constant migration of vortices away from the rotation axis. 
The angular momentum reservoir builds when a large number of vortices is unable to migrate because of strong pining to nuclei in the crust, thereby preventing that superfluid component from slowing down \citep{ai75,aaps84}.
A glitch happens when millions of vortices catastrophically unpin and are allowed to move outwards, transferring angular momentum to the crust and all the other sections of the star in corotation (the \emph{charged} component).
Because of the sudden increase of spin frequency, another superfluid component will become decoupled from the general rotation and will re-couple to the \emph{charged} component on time scales which depend on the levels of friction between them \citep{,hps12,nbh15}. 
The observed recoveries are the product of the re-coupling of this superfluid component. 
The interaction between the superfluid and the \emph{charged} component is known as \emph{mutual friction} \citep[e.g.][]{asc06} and can be modelled with one parameter, which can take different values across the star 
and reproduce different recovery timescales after glitches \citep{ha14}.
Therefore, glitch recoveries can be seen as the (slow) response of a superfluid component to the spin up of the star.

The large inter-glitch braking indices observed between glitches could be caused by the progressive coupling of this superfluid to the rotation of the \emph{charged} component and the consequent increase of the effective moment of inertia (i.e. $\dot{I}>0$, produced by $n_\text{obs}=n_\text{nig}$ and $n_\text{nig}>3$ in Eq. \ref{idot}). 
Because of the exponential evolution of \nudot\ after glitches, and the gradual reduction of \nuddot\ (Fig. \ref{allf1}), the braking indices between glitches evolve from very large values and converge to $n_\text{ig}\geq 10$. 
The fact that the braking indices between glitches never reach values close to, or below, $3$ suggests that equilibrium is never reached and the re-coupling is unfinished by the time a new glitch happens.
Maybe, because glitches occur before the star is fully relaxed and every new glitch imposes new, different rotation rates, there is a progressive decoupling of superfluid residing in areas where mutual friction is low.
This was already discussed by \citet{smi99} for the Vela pulsar and it is discussed  in detail by \citet{aeka15} for the case of PSR J0537$-$6910, the pulsar with the highest rate of large glitches and the smallest braking index ($n=-1.2$).

The cumulative effect over many glitches would imply a net long-term decrease of the effective moment of inertia, with timescales of a few $10^4$\,yr,  producing the observed $n\sim2$.
\citet{smi99} argues that this process cannot be sustained for a long time if the superfluid involved corresponds to the $\sim 2\%$ that is believed to participate in the glitches \citep{lel99} and that the observed low braking index is more likely to be caused by an increase of the magnetic moment.
Similar limitations are found if the decoupling of the superfluid is driven by the pining of vortices in the core of the star \citep{ha12} where, to reproduce some observed braking indices, fractions as large as 20-30 \% of the total moment of inertia would have to be pinned.\footnote{Private communication with Elena Kantor.}
Indeed, the fraction of superfluid that would have to be decoupled in the Vela pulsar after $1$\,kyr, to produce $n=1.7$, amounts to $\sim5\%$ of the total.
It is unclear to us whether this amount of superfluid (or more) is available in the regions where mutual friction is low.

We note that probably not all low braking indices are driven by glitch activity and that other mechanisms must be operating. 
The high magnetic field pulsar PSR J1734$-$3333 has exhibited only one small  glitch\footnote{\url{http://www.jb.man.ac.uk/pulsar/glitches.html}} in more than $15$ years of observation and there are no signs of recent large glitches prior to the observations ($n_\text{ig}=0.9\pm0.2$; Table \ref{tbl:enes} and \citet{elk+11}).   
Thus, its spin evolution is likely dominated by the underlying braking mechanism, which has been proposed to be driven by either a fall back disk \citep{cea+13,cl16} or an increasing dipole magnetic field \citep{elk+11}. 
The increase of the dipole component of the magnetic field could be driven by the evolution (via ohmic difussion or Hall drift) of a stronger magnetic field  underneath the stellar surface.
The presence of very strong internal magnetic fields inside neutron stars
has already been used to explain not only the low braking indices of the classical young pulsars \citep{smi99,lyn04,elk+11,gc15}, but also other phenomena among other young neutron stars, like CCOs  \citep{ho11,vp12}, magnetars \citep{mpm15,gpr+15} and the missing neutron star in the SNR of SN 1987A \citep{gpz99}. 
\citet{ho2015} relates glitch activity with magnetic field evolution and proposes a causal connection between them, which could explain $n\sim2$ among glitching pulsars.
Whether the two phenomena are connected or not is still unclear, but it seems plausible that the observed long-term evolution of young pulsars is controlled by magnetic field evolution, which is at some point accompanied (perhaps overriden, while glitch activity is high) by a variable (decreasing) moment of inertia driven by the decoupling of  superfluid.

\section{Conclusions}
We have analysed the \nudot\ evolution of the Vela pulsar and the Vela-like pulsars \bb\ and \cc, applying a newly devised method to measure their long-term braking indices, $n$.
The new method is based on the same principles as the one created by \citet[][for the Vela pulsar]{lpgc96} but it is less dependent on free parameters.
Vela and the other two pulsars exhibit a regular glitch behaviour, with glitches of similar sizes occurring at semi-regular time intervals and followed by recoveries of similar morphology.
The evolutions of other six, less regular young glitching pulsars were also analysed and we offer new long-term braking index estimates for four of them.
All these glitching pulsars exhibit values $n\leq2$.

Regardless of the mechanisms behind the observed $n$ values, it seems clear  that there is a population of young glitching pulsars evolving with low long-term braking indices, or moving with slopes $\geq0$ across the \ppdd.
The time these pulsars spend evolving this way may not be negligible.
In that case, these upward movements imply a new flow of pulsars on the diagram that should be incorporated in evolution models like those used for population syntheses.
Furthermore, using the real braking index values (rather than $n=3$), the characteristic ages of these pulsars can become considerably larger.
We briefly discuss the case of PSR \aaa\ and conclude that with a low braking index, such as the one we obtain, the pulsar could be as old as is necessary for the association with SNR G$5.4-1.2$ to be real.

The effects of large glitches, in the form of large \nudot\ jumps followed by exponential-like recoveries and large inter-glitch \nuddot\ values (conducive to large inter-glitch braking indices $n_\text{ig}>10$), fully shape the short-term spin evolution of these pulsars and might also be responsible of their long-term evolution. 
The low long-term braking indices could be the result of a decreasing effective moment of inertia caused by the progressive decoupling of superfluid at the glitches.
This process could be superimposed on the effects of magnetic field evolution, which in the case of dipole field growth would manifest as low braking indices too.

\section*{Acknowledgements}
Pulsar research at JBCA is supported by a Consolidated Grant from the UK Science and Technology Facilities Council (STFC). 
This paper includes archived data from the Parkes radio telescope obtained through the Australia Telescope Online Archive and the CSIRO Data Access Portal (http://data.csiro.au).  The Parkes radio telescope is part of the Australia Telescope National Facility which is funded by the Commonwealth of Australia for operation as a National Facility managed by CSIRO.
The authors thank Wynn C.~G. Ho for useful comments.
CME acknowledges support from {\sc {CONICYT}} ({\sc FONDECYT} postdoctorado 3130512 and PIA ACT1405).



\bibliographystyle{mnras}
\bibliography{journals,modrefs,psrrefs,crossrefs,cme}





\bsp	
\label{lastpage}
\end{document}